# Thermocapillary effect on the cross-stream migration of a surfactant-laden droplet in Poiseuille flow

Sayan Das, Shubhadeep Mandal, S K Som and Suman Chakraborty†

Department of Mechanical Engineering, Indian Institute of Technology Kharagpur,
Kharagpur – 721302, India

The motion of a viscous droplet in unbounded Poiseuille flow under the combined influence of bulk-insoluble surfactant and linearly varying temperature field aligned in the direction of imposed flow is studied analytically. Neglecting fluid inertia, thermal convection and shape deformation, asymptotic analysis is performed to obtain the velocity of a force-free surfactant-laden droplet. The droplet speed and direction of motion are strongly influenced by the interfacial transport of surfactant which is governed by surface Péclet number. The present study is focused on the following two limiting situations of surfactant transport: (*i*) surface diffusion dominated surfactant transport considering small surface Péclet number, and (*ii*) surface convection governed surfactant transport considering high surface Péclet number. Thermocapillary-induced Marangoni stress, strength of which relative to viscous stress is represented by thermal Marangoni number, has strong influence on the distribution of surfactant on the droplet surface. Temperature field not only affects the axial velocity of the droplet but also has significant effect on the cross-stream velocity of the droplet in spite of the fact that the temperature gradient is aligned with the Poiseuille flow direction. When the imposed temperature increases in the direction of Poiseuille flow, the droplet migrates towards the flow centerline. The magnitude of both axial and cross-stream velocity components increases with the thermal Marangoni number. However, when the imposed temperature decreases in the direction of Poiseuille flow, the magnitude of both axial and cross-stream velocity components may increase or decrease with the thermal Marangoni number. Most interestingly, the droplet moves either towards the flow centerline or away from it. Present study shows a critical value of the thermal Marangoni number beyond which the droplet moves away from the flow centerline which is in sharp contrast to the motion of a surfactant-laden droplet in isothermal flow for which droplet always moves towards the flow centerline.

†Email address for correspondence: suman@mech.iitkgp.ernet.in

## 1. Introduction

The dynamics of suspending droplets has been a recent topic of interest to science and engineering community due to its wide application in microfluidic devices (Baroud et al. 2010; Seemann et al. 2012; Stone et al. 2004). In such devices, droplets are used in analytic detection, reagent mixing, drug delivery and cell encapsulation process to name a few (Teh et al. 2008; Huebner et al. 2008; Di Carlo et al. 2007; Baroud et al. 2010; Zhu & Fang 2013). Surfactants (or surface-active agents) are common in these droplet-based devices. Surfactants are used as additives in droplet-based systems to enhance droplet generation and stability of emulsion (Baret 2012). Presence of surfactants at the droplet interface not only lowers the interfacial tension but also can induce a Marangoni stress if spatial variation of surfactants is nonuniform (Leal 2007). So, to perform optimal functionalities of the concerned devices, the fundamental understanding of the motion of surfactant-laden droplet is of prime importance.

Several theoretical and experimental works have been performed to study the motion of droplets in Poiseuille flow (Stan et al. 2011; Chen et al. 2014; Leal 1980). In the creeping flow limit, a non-deformable, spherical, Newtonian droplet suspended in another Newtonian fluid with clean fluid-fluid interface moves in the direction of flow in an unbounded Poiseuille flow (Hetsroni & Haber 1970). This can be well explained by the symmetry-under-flow-reversal argument which is valid for linear governing equations and boundary conditions (Leal 2007). Interesting things happen in the presence of nonlinear effects such as deformation, viscoelasticity and inertia. A deformable droplet not only moves in the direction of flow, but also moves in the cross-stream direction (Haber & Hetsroni 1971; Wohl & Rubinow 1974; Chan & Leal 1979; Mortazavi & Tryggvason 2000; Griggs et al. 2007; Mandal et al. 2015). Chan & Leal (1979) have found that a deformable droplet moves towards the channel centerline when $\lambda < 0.5$ or $\lambda > 10$ (where $\lambda$ is the droplet to medium viscosity ratio), while droplet moves away from the centerline when $0.5 < \lambda < 10$. Several studies have also reported cross-stream migration of non-deformable droplet in Poiseuille flow in the presence of fluid inertia (Karnis et al. 1966; Hur et al. 2011; Chen et al. 2014) or fluid viscoelasticity (Chan & Leal 1979; Mukherjee & Sarkar 2013; Mukherjee & Sarkar 2014). These deformation-, inertia- and viscoelasticity-induced cross-stream migration characteristics are used in microfluidic devices as sorting mechanisms (Sajeesh & Sen 2014; Amini et al. 2014; Hatch et al. 2013). Very recently, Hanna & Vlahovska (2010) and Pak et al. (2014) have found that surfactant-induced Marangoni stress at the droplet interface can induce a cross-stream migration of a spherical droplet in Poiseuille flow even in the absence of deformation, inertia and viscoelasticity.

Studies on the droplet motion in the presence of external effects such as electric (Ahn et al. 2006; Link et al. 2006; Bandopadhyay et al. 2016; Mandal et al. 2016), magnetic (Seemann et al. 2012), temperature (Karbalaei et al. 2016) and acoustic fields (Seemann et al. 2012) are gaining much importance nowadays due to the ease with which these fields can be applied in respective applications. The presence of these fields induces an imbalance in stresses at the droplet interface and modifies the net force acting on the droplet which in turn alters the droplet velocity and associated flow field. Out of all the effects mentioned above, in the present study we focus on the effect of temperature field. The variation of temperature field across the droplet induces Marangoni stress at the droplet interface which induces droplet motion even in the absence of imposed flow (Young et al. 1959). Following Young et al. (1959), several studies have considered the thermocapillary motion of droplet in quiescent medium to study the following aspects: droplet deformation (Nadim et al. 1990), fluid inertia (Haj-Hariri et al. 1990), thermal convection (Balasubramaniam & Subramanian 2004; Yariv & Shusser 2006; Zhang et al. 2001) and bounding wall (Meyyappan & Subramanian 1987; Barton & Shankar Subramanian 1990; Barton & Subramanian 1991; Chen 1999; Chen 2003). In recent studies, Raja Sekhar and co-works have investigated the effect of thermocapillary-induced Marangoni stress on the droplet velocity in the presence of imposed flow field (Choudhuri & Raja Sekhar 2013; Sharanya & Raja Sekhar 2015). In the absence of shape deformation, surfactants and fluid inertia, their studies show that the effects of temperature field and imposed background flow can be linearly combined to obtain the final droplet velocity.

In a recent work, we have investigated the axisymmetric motion of a surfactant-laden droplet in combined presence of linearly varying temperature field and imposed Poiseuille flow (Das et al. 2016). However, there is no study present in the literature which investigates the combined effect of temperature and imposed Poiseuille flow on the cross-stream migration characteristics of a droplet in the presence of bulk-insoluble surfactants. In the present study, we analytically obtain the droplet velocity in an unbounded Poiseuille flow considering both thermocapillary-induced and surfactant-induced Marangoni stresses at the droplet interface. We show that in the presence of nonuniform surfactant distribution at the droplet interface, the effects of temperature field and imposed background flow cannot be linearly combined to obtain the final droplet velocity. By neglecting fluid inertia, thermal convection and shape deformation, we perform asymptotic analysis for two different limits: (i) when the surface diffusion of the surfactants dominates interfacial transport, and (ii) when the surface convection of the surfactants dominates interfacial transport. Most important finding from the present study is that the thermocapillary effect not only alters the magnitude of droplet speed but also has the ability to change the direction of cross-stream migration of the droplet.

## 2. Problem formulation

### 2.1. *System description*

Present system consists of a spherical, neutrally buoyant, Newtonian droplet (density $\rho$, viscosity $\mu_i$, and thermal conductivity $k_i$) of radius $a$ suspended in another Newtonian medium (density $\rho$, viscosity $\mu_e$, and thermal conductivity $k_e$). Bulk-insoluble surfactants are present at the droplet interface. All the material properties are assumed to be constant, except the interfacial tension $\left(\overline{\sigma}\right)$. The interfacial tension depends on the interface temperature $\left(\overline{T}_s\right)$ and the local surfactant concentration $\left(\overline{\Gamma}\right)$ along the interface of the droplet. In a quiescent medium, the surfactants are uniformly distributed over the droplet surface. The concentration of surfactant at equilibrium is denoted by $\overline{\Gamma}_{eq}$ and corresponding interfacial tension is denoted by $\overline{\sigma}_{eq}$. This equilibrium is disturbed by application of background Poiseuille flow $\left(\overline{\mathbf{V}}_\infty\right)$ and linearly varying temperature field $\left(\overline{T}_\infty\right)$. Imposed Poiseuille flow alters the interfacial tension via the convection of surfactants at the droplet interface. On the other hand, the effect of temperature field on the interfacial tension is twofold: (i) it directly alters the interfacial tension as interfacial tension is a function of temperature at the droplet interface, and (ii) thermocapillary-induced Marangoni stress induces transport of surfactants at the droplet interface. So, combined presence of Poiseuille flow and temperature field leads to generation of Marangoni stresses which will affect the droplet velocity $\left(\overline{\mathbf{U}}\right)$. The main objective of the present study is to investigate the effect of the Marangoni stresses on the droplet velocity. Towards this, we consider a spherical coordinate system $\left(\overline{r},\theta,\varphi\right)$ which is attached at the droplet centroid (refer to figure 1).

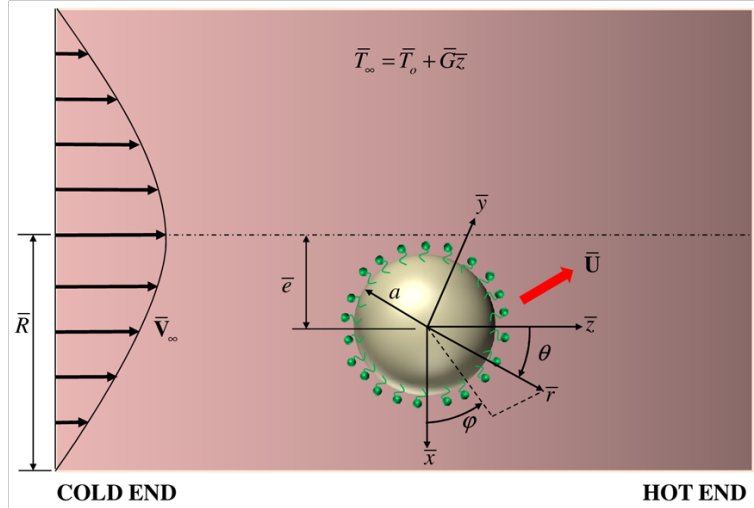

FIGURE 1. Schematic of a surfactant-laden droplet of radius *a* suspended in a cylindrical Poiseuille flow $\left(\overline{\mathbf{V}}_\infty\right)$. The imposed temperature field $\left(\overline{T}_\infty\right)$ varies linearly in the axial direction ($\overline{z}$-direction). The droplet is placed at an eccentric location of distance $\overline{e}$ from the channel centerline. Both spherical $\left(r,\theta,\varphi\right)$ and Cartesian $\left(\overline{x},\overline{y},\overline{z}\right)$ coordinates are shown. Important to note that the $\overline{x}$ axis is always directed outward from the flow centerline.

## 2.2. *Assumptions*

Major assumptions made in this study to simplify the governing equations and the boundary conditions are as follows: (i) The thermal problem is governed by the conduction of heat which is based on the fact that the thermal Péclet number ($Pe_T = \overline{V}_c a / \alpha_e$, where $\alpha_e$ is the thermal diffusivity of the suspending medium and $\overline{V}_c$ is the centerline velocity of imposed Poiseuille flow) is small enough so that the advective transport of thermal energy is negligible. This consideration makes the thermal problem linear. (ii) The flow problem is governed by the viscous and pressure forces. This is based on the fact that the Reynolds number $\left( Re = \rho \overline{V}_c a / \mu_e \right)$ is small enough so that the fluid inertia is negligible. (iii) The droplet shape is considered as a perfect sphere. This is based on the fact that capillary number $\left( Ca = \mu_e \overline{V}_c / \overline{\sigma}_o \right)$ is small enough so that the interfacial tension dominates over the viscous stresses. The typical values of these non-dimensional numbers can be obtained from the existing experimental data (Nallani & Subramanian 1993). Nallani & Subramanian (1993) have performed experiments on the thermocapillary motion of a methanol droplet of radius $50\,\mu\text{m}$ suspended in silicone oil (with $\rho = 955\,\text{kg/m}^3$, $\mu_e = 0.0478\,\text{Ns/m}^2$, $k_e = 0.1\,\text{W/mK}$ and $c_{pe} = 1800\,\text{J/kgK}$). Considering $\overline{V}_c = 10^{-4}\,\text{m/s}$ (which is common in microfluidic devices) and $\overline{\sigma}_o = 10^{-3}\,\text{N/m}$, we obtain $Pe_T \sim 0.01$, $Re \sim 10^{-4}$ and $Ca \sim 0.001$. So the consideration of negligible thermal convection, fluid inertia and shape deformation is justified. (iv) The surfactants are insoluble in the bulk fluid (Baret 2012). (v) The surfactant distribution along the interface of the droplet does not affect the heat transfer process (Kim & R. Shankar Subramanian 1989). (vi) The interfacial tension is linearly dependent on the temperature and the surfactant concentration at the interface of the droplet (Homsy & Meiburg 1984; Carpenter & Homsy 1985). (vii) We consider the imposed Poiseuille flow to be a unbounded one which disregards the effects of bounding walls.

## 2.3. *Governing equations and boundary conditions*

As the thermal Péclet number is small, the thermal problem is governed by the Laplace equation of the form

$$\left. \begin{array}{l} \overline{\nabla}^2 \overline{T}_i = 0, \\ \overline{\nabla}^2 \overline{T}_e = 0, \end{array} \right\} \tag{1}$$

where $\overline{T}_i$ and $\overline{T}_e$ represent the temperature field inside and outside the droplet. Subscripts '$i$' and '$e$' are used to represent quantities inside and outside the droplet, respectively. The temperature field outside the droplet $\left( \overline{T}_e \right)$ satisfies the far-field condition

$$\text{as } \overline{r} \to \infty, \quad \overline{T}_e = \overline{T}_\infty, \tag{2}$$

where $\overline{T}_\infty$ with respect to a spherical coordinate system (origin at the droplet centroid) is represented as

$$\overline{T}_\infty = \overline{T}_o + \overline{G}\overline{z},$$  (3)

where $\overline{G}$ is the temperature gradient along $\overline{z}$ and $\overline{T}_o$ is the reference temperature. In the present study, we consider variation of $\overline{T}_\infty$ along the direction of imposed Poiseuille flow direction. The imposed temperature increases (or decreases) in the direction of imposed Poiseuille flow for $\overline{G} > 0$ (or $\overline{G} < 0$). The temperature field inside the droplet $(\overline{T}_i)$ is bounded at the centre of the droplet $(\overline{r} = 0)$. In addition to these, $\overline{T}_i$ and $\overline{T}_e$ satisfy the temperature continuity and heat flux continuity at the droplet surface $(\overline{r} = a)$:

$$\left.\begin{array}{ll} \text{at } \overline{r} = a, & \overline{T}_i = \overline{T}_e, \\ \text{at } \overline{r} = a, & k_i \dfrac{\partial \overline{T}_i}{\partial \overline{r}} = k_e \dfrac{\partial \overline{T}_e}{\partial \overline{r}}. \end{array}\right\}$$  (4)

As the Reynolds number is small, the flow problem is governed by the Stokes and continuity equations in the following form

$$\left.\begin{array}{l} -\overline{\nabla}\overline{p}_i + \mu_i \overline{\nabla}^2 \overline{\mathbf{u}}_i = \mathbf{0}, \ \ \overline{\nabla} \cdot \overline{\mathbf{u}}_i = 0, \\ -\overline{\nabla}\overline{p}_e + \mu_e \overline{\nabla}^2 \overline{\mathbf{u}}_e = \mathbf{0}, \ \ \overline{\nabla} \cdot \overline{\mathbf{u}}_e = 0, \end{array}\right\}$$  (5)

where $\overline{\mathbf{u}}$ and $\overline{p}$ represent the velocity and pressure fields. The velocity and pressure fields outside the droplet satisfy the far-field condition

$$\left.\begin{array}{ll} \text{as } \overline{r} \to \infty, & \overline{\mathbf{u}}_e = \overline{\mathbf{V}}_\infty - \overline{\mathbf{U}}, \\ \text{as } \overline{r} \to \infty, & \overline{p}_e = \overline{p}_\infty, \end{array}\right\}$$  (6)

where $\overline{p}_\infty$ is the pressure field associated with $\overline{\mathbf{V}}_\infty$. The imposed Poiseuille flow $(\overline{\mathbf{V}}_\infty)$ with respect to a spherical coordinate system (origin at the droplet centroid) is represented as

$$\overline{\mathbf{V}}_\infty = \overline{V}_c \left[ 1 - \frac{\overline{e}^2}{\overline{R}^2} - \frac{\overline{r}^2}{\overline{R}^2}\sin^2\theta - \frac{2\overline{r}}{\overline{R}^2}\overline{e}\cos\varphi\sin\theta \right] \mathbf{e}_z,$$  (7)

where $\overline{e}$ and $\overline{R}$ are the position of the droplet centroid and the location of zero imposed velocity, which are measured from the centerline of the flow. The velocity and pressure fields inside the droplet $(\overline{\mathbf{u}}_i, \overline{p}_i)$ are bounded at the centre of the droplet $(\overline{r} = 0)$. In addition to these, the velocity and pressure fields satisfy the following interfacial conditions at the droplet surface $(\overline{r} = a)$:

$$\left.\begin{array}{ll}
\text{at } \overline{r}=a, & \overline{\mathbf{u}}_i=\overline{\mathbf{u}}_e, \\
\text{at } \overline{r}=a, & \overline{\mathbf{u}}_i\cdot\mathbf{e}_r=\overline{\mathbf{u}}_e\cdot\mathbf{e}_r=0, \\
\text{at } \overline{r}=a, & \left(\overline{\boldsymbol{\tau}}_e\cdot\mathbf{e}_r-\overline{\boldsymbol{\tau}}_i\cdot\mathbf{e}_r\right)\cdot\left(\mathbf{I}-\mathbf{e}_r\mathbf{e}_r\right)=-\overline{\boldsymbol{\nabla}}_s\overline{\sigma},
\end{array}\right\} \tag{8}$$

where $\overline{\boldsymbol{\tau}}_i=-\overline{p}_i\mathbf{I}+\mu_i\left[\overline{\boldsymbol{\nabla}}\overline{\mathbf{u}}_i+\left(\overline{\boldsymbol{\nabla}}\overline{\mathbf{u}}_i\right)^T\right]$ and $\overline{\boldsymbol{\tau}}_e=-\overline{p}_e\mathbf{I}+\mu_e\left[\overline{\boldsymbol{\nabla}}\overline{\mathbf{u}}_e+\left(\overline{\boldsymbol{\nabla}}\overline{\mathbf{u}}_e\right)^T\right]$ represent the hydrodynamic stress tensors inside and outside the droplet, and $\overline{\boldsymbol{\nabla}}_s=\left[\overline{\boldsymbol{\nabla}}-\mathbf{e}_r\left(\mathbf{e}_r\cdot\overline{\boldsymbol{\nabla}}\right)\right]$ represents the surface gradient operator. The first condition represents the continuity of velocity field, the second condition represents the kinematic condition, and the third condition represents the balance between tangential component of hydrodynamic and Marangoni stresses. The normal stress balance is not relevant to the present study as we are considering spherical droplet shape (Choudhuri & Raja Sekhar 2013).

The Marangoni stress depends on the variation of interfacial tension $\left(\overline{\sigma}\right)$. Assuming linear relationship, the dependence of interfacial tension on the temperature and surfactant concentration can be represented as (Kim & R. Shankar Subramanian 1989; Homsy & Meiburg 1984; Carpenter & Homsy 1985)

$$\overline{\sigma}=\overline{\sigma}_o-\beta\left(\overline{T}_s-\overline{T}_o\right)-R_g\overline{T}_o\overline{\Gamma}, \tag{9}$$

where $\overline{\sigma}_o$ is the interfacial tension at the reference temperature $\overline{T}_o$ in absence of any surfactant, $\overline{T}_s=\overline{T}_i\big|_{\overline{r}=a}$ is the temperature at the droplet surface, $R_g$ is the ideal gas constant, and $\beta=-d\overline{\sigma}/d\overline{T}_s$. The surfactant concentration $\left(\overline{\Gamma}\right)$ at the droplet surface is governed by the following convection-diffusion equation (Stone & Leal 1990; Stone 1990):

$$\overline{\boldsymbol{\nabla}}_s\cdot\left(\overline{\mathbf{u}}_s\overline{\Gamma}\right)=D_s\overline{\nabla}_s^2\overline{\Gamma}, \tag{10}$$

where $D_s$ denotes the surface diffusivity of the surfactants, and $\overline{\mathbf{u}}_s$ denotes the fluid velocity on the droplet surface.

### 2.4. *Non-dimensional form of governing equations and boundary conditions*

Now, we apply the following non-dimensional scheme to the above set of governing equations and boundary conditions to obtain the relevant dimensionless form: $r=\overline{r}/a$, $\mathbf{u}=\overline{\mathbf{u}}/V_c$, $T=\left(\overline{T}-\overline{T}_o\right)\big/\left|\overline{G}\right|a$, $\Gamma=\overline{\Gamma}/\overline{\Gamma}_{eq}$, $\sigma=\overline{\sigma}/\overline{\sigma}_o$, $p=\overline{p}\big/\left(\mu_e\overline{V}_c/a\right)$, and $\boldsymbol{\tau}=\overline{\boldsymbol{\tau}}\big/\left(\mu_e\overline{V}_c/a\right)$. The various material ratios that appear are: viscosity ratio $\lambda=\mu_i/\mu_e$ and conductivity ratio $\delta=k_i/k_e$. On the other hand, different non-dimensional numbers that appear are: surface Péclet number $Pe_s = \overline{V}_c a / D_s$ (the ratio of the strength of the advection of the surfactant as compared to the diffusion of the same at the droplet interface), surfactant Marangoni number $Ma_\Gamma = \overline{\Gamma}_{eq} R_g \overline{T}_o / \mu_e \overline{V}_c$ (the ratio of surfactant-induced Marangoni stress and viscous stress), and thermal Marangoni number $Ma_T = \beta |\overline{G}| a / \mu_e \overline{V}_c$ (the ratio of thermocapillary-induced Marangoni stress and viscous stress).

Imposing the above mentioned scales, we obtained the non-dimensional form of the Laplace equation which governs the thermal problem as

$$\left.\begin{aligned} \nabla^2 T_i = 0, \\ \nabla^2 T_e = 0, \end{aligned}\right\} \tag{11}$$

which are subjected to the following boundary conditions

$$\left.\begin{aligned} &\text{as } r \to \infty, \, T_e = \zeta r \cos\theta, \\ &T_i \text{ is bounded at } r = 0, \\ &\text{at } r = 1, \ T_i = T_e, \\ &\text{at } r = 1, \ \delta \frac{\partial T_i}{\partial r} = \frac{\partial T_e}{\partial r}. \end{aligned}\right\} \tag{12}$$

The factor $\zeta = \overline{G} / |\overline{G}|$ signifies the direction of imposed temperature field. The imposed temperature increases (or decreases) in the direction of imposed Poiseuille flow for $\zeta = 1$ (or $\zeta = -1$). The flow problem is governed by the following non-dimensional equations

$$\left.\begin{aligned} -\nabla p_i + \lambda \nabla^2 \mathbf{u}_i = \mathbf{0}, \ \nabla \cdot \mathbf{u}_i = 0, \\ -\nabla p_e + \nabla^2 \mathbf{u}_e = \mathbf{0}, \ \nabla \cdot \mathbf{u}_e = 0, \end{aligned}\right\} \tag{13}$$

which are subjected to the following boundary conditions

$$\left.\begin{aligned} &\text{at } r \to \infty, \ \left(\mathbf{u}_e, p_e\right) = \left(\mathbf{V}_\infty - \mathbf{U}, p_\infty\right), \\ &\mathbf{u}_i \text{ is bounded at } r = 0, \\ &\text{at } r = 1, \ \mathbf{u}_i \cdot \mathbf{e}_r = \mathbf{u}_e \cdot \mathbf{e}_r = 0, \\ &\text{at } r = 1, \ \mathbf{u}_i = \mathbf{u}_e, \\ &\text{at } r = 1, \ \left(\boldsymbol{\tau}_e \cdot \mathbf{e}_r - \boldsymbol{\tau}_i \cdot \mathbf{e}_r\right) \cdot \left(\mathbf{I} - \mathbf{e}_r \mathbf{e}_r\right) = Ma_\Gamma \nabla_s \Gamma + Ma_T \nabla_s T_s. \end{aligned}\right\} \tag{14}$$

The last boundary condition (tangential stress balance) is obtained by substituting the dimensionless form of interfacial tension as

$$\sigma = 1 - Ca\left(Ma_T T + Ma_\Gamma \Gamma\right).$$ (15)

The surfactant transport equation, when non-dimensionalized, takes the form

$$Pe_s \nabla_s \cdot \left(\mathbf{u}_s \Gamma\right) = \nabla_s^2 \Gamma.$$ (16)

In addition to this, the surfactant concentration should also satisfy the following constraint to conserve the total mass of the surfactants on the droplet surface in the following form

$$\int_{\varphi=0}^{2\pi} \int_{\theta=0}^{\pi} \Gamma\left(\theta, \varphi\right) \sin\theta \, d\theta \, d\varphi = 4\pi.$$ (17)

A closer look into equation (16) reveals the nonlinearity in the surfactant convection term. This makes the flow and surfactant transport coupled due to which it is impossible to obtain analytical solution for arbitrary values of $Pe_s$. Towards making analytical treatment, we focus on the following two regimes (Pak et al. 2014; Hanna & Vlahovska 2010): (i) $Pe_s \ll 1$ which signifies that the surface diffusion dominates the surfactant transport, and (ii) $Pe_s \to \infty$ which signifies that the surface convection governs the surfactant transport. First limit represents the physical situation of large surface diffusivity of surfactants, while the second limit represents the physical situation of negligible surface diffusivity of surfactants (Vlahovska et al. 2009; Vlahovska et al. 2005; Kim & R.Shankar Subramanian 1989).

## 3. Asymptotic solution

### 3.1. *Representation of field variables in terms of spherical harmonics*

As the temperature fields inside and outside the droplet satisfy the Laplace equation, we can represent them in terms of spherical solid harmonics as (Choudhuri & Raja Sekhar 2013)

$$\left.\begin{aligned} T_i &= \sum_{n=0}^{\infty} \sum_{m=0}^{n} \left[a_{n,m} r^n \cos\left(m\varphi\right) + \hat{a}_{n,m} r^n \sin\left(m\varphi\right)\right] P_{n,m}\left(\cos\theta\right), \\ T_e &= \zeta \, r P_{1,0}\left(\cos\theta\right) + \sum_{n=0}^{\infty} \sum_{m=0}^{n} \left[b_{-n-1,m} r^{-n-1} \cos\left(m\varphi\right) + \hat{b}_{-n-1,m} r^{-n-1} \sin\left(m\varphi\right)\right] P_{n,m}\left(\cos\theta\right), \end{aligned}\right\}$$ (18)

where $P_{n,m}\left(\cos\theta\right)$ is the associated Legendre polynomial of degree $n$ and order $m$. The temperature field inside the droplet $\left(T_i\right)$ is constructed by linearly combining the growing spherical solid harmonics which automatically satisfies the boundedness of $T_i$ at $r = 0$. On the other hand, the temperature field outside the droplet $\left(T_e\right)$ is constructed by linearly combining the far-field temperature field $T_\infty = \zeta r P_{1,0}\left(\cos\theta\right)$ and the decaying spherical solid harmonics which automatically satisfies the boundedness of $T_e$ as $r \to \infty$. In equation (18) $a_{n,m}$, $\hat{a}_{n,m}$, $b_{-n-1,m}$ and $\hat{b}_{-n-1,m}$ are the unknown coefficients of the spherical solid harmonics which will be obtained by using rest of the boundary conditions (continuity of temperature and heat flux across the droplet interface). The surface temperature $\left(T_s\right)$ can be expressed as

$$T_s = \sum_{n=0}^{\infty}\sum_{m=0}^{n}\left[T_{n,m}\cos\left(m\varphi\right) + \hat{T}_{n,m}\sin\left(m\varphi\right)\right]P_{n,m}\left(\cos\theta\right), \tag{19}$$

where $T_{n,m}$ and $\hat{T}_{n,m}$ are the coefficients of the spherical surface harmonics. Similarly, the surfactant concentration $\Gamma$ can be represented in terms of spherical surface harmonics as (Pak et al. 2014)

$$\Gamma = \sum_{n=0}^{\infty}\sum_{m=0}^{n}\left[\Gamma_{n,m}\cos\left(m\varphi\right) + \hat{\Gamma}_{n,m}\sin\left(m\varphi\right)\right]P_{n,m}\left(\cos\theta\right), \tag{20}$$

where $\Gamma_{n,m}$ and $\hat{\Gamma}_{n,m}$ will be obtained by solving the surfactant transport equation.

As the velocity and pressure fields inside the droplet satisfy the Stokes and continuity equations, we can represent them in terms of the growing spherical solid harmonics using the Lamb's general solution as (Haber & Hetsroni 1972)

$$\left.\begin{aligned} \mathbf{u}_i &= \sum_{n=1}^{\infty}\left[\nabla\times\left(\mathbf{r}\chi_n\right) + \nabla\Phi_n + \frac{n+3}{2\left(n+1\right)\left(2n+3\right)\lambda}r^2\nabla p_n - \frac{n}{\left(n+1\right)\left(2n+3\right)\lambda}\mathbf{r}p_n\right], \\ p_i &= \sum_{n=0}^{\infty}p_n, \end{aligned}\right\} \tag{21}$$

where $p_n$, $\Phi_n$ and $\chi_n$ are the growing spherical solid harmonics of the form

$$\left.\begin{aligned} p_n &= \lambda r^n \sum_{m=0}^{n}\left[A_{n,m}\cos\left(m\varphi\right) + \hat{A}_{n,m}\sin\left(m\varphi\right)\right]P_{n,m}\left(\cos\theta\right), \\ \Phi_n &= r^n \sum_{m=0}^{n}\left[B_{n,m}\cos\left(m\varphi\right) + \hat{B}_{n,m}\sin\left(m\varphi\right)\right]P_{n,m}\left(\cos\theta\right), \\ \chi_n &= r^n \sum_{m=0}^{n}\left[C_{n,m}\cos\left(m\varphi\right) + \hat{C}_{n,m}\sin\left(m\varphi\right)\right]P_{n,m}\left(\cos\theta\right). \end{aligned}\right\} \tag{22}$$

Similarly, the velocity and pressure fields outside the droplet can be represented in terms of the far-field velocity field and the decaying spherical solid harmonics using the Lamb's solution as (Haber & Hetsroni 1972)

$$\left.\begin{aligned}\mathbf{u}_e &= \left(\mathbf{V}_\infty - \mathbf{U}\right) + \sum_{n=1}^{\infty}\left[\nabla\times\left(\mathbf{r}\chi_{-n-1}\right) + \nabla\Phi_{-n-1} - \frac{n-2}{2n\left(2n-1\right)}r^2\nabla p_{-n-1} + \frac{n+1}{n\left(2n-1\right)}\mathbf{r}p_{-n-1}\right],\\ p_e &= p_\infty + \sum_{n=0}^{\infty}p_{-n-1},\end{aligned}\right\} \quad (23)$$

where $p_{-n-1}$, $\Phi_{-n-1}$ and $\chi_{-n-1}$ are the decaying spherical solid harmonics of the form

$$\left.\begin{aligned}p_{-n-1} &= r^{-n-1}\sum_{m=0}^{n}\left[A_{-n-1,m}\cos\left(m\varphi\right) + \hat{A}_{-n-1,m}\sin\left(m\varphi\right)\right]P_{n,m}\left(\cos\theta\right),\\ \Phi_{-n-1} &= r^{-n-1}\sum_{m=0}^{n}\left[B_{-n-1,m}\cos\left(m\varphi\right) + \hat{B}_{-n-1,m}\sin\left(m\varphi\right)\right]P_{n,m}\left(\cos\theta\right),\\ \chi_{-n-1} &= r^{-n-1}\sum_{m=0}^{n}\left[C_{-n-1,m}\cos\left(m\varphi\right) + \hat{C}_{-n-1,m}\sin\left(m\varphi\right)\right]P_{n,m}\left(\cos\theta\right).\end{aligned}\right\} \quad (24)$$

The unknown coefficients $A_{n,m}$, $B_{n,m}$, $C_{n,m}$, $\hat{A}_{n,m}$, $\hat{B}_{n,m}$, $\hat{C}_{n,m}$, $A_{-n-1,m}$, $B_{-n-1,m}$, $C_{-n-1,m}$, $\hat{A}_{-n-1,m}$, $\hat{B}_{-n-1,m}$ and $\hat{C}_{-n-1,m}$ will be obtained by using rest of the boundary conditions (normal velocity, tangential velocity and tangential stress conditions).

### 3.2. *Solution for* $Pe_s \ll 1$

Irrespective of the value of $Pe_s$, the thermal problem is independent of the flow field and surfactant concentration. So, we obtain the solution for temperature field inside and outside the droplet in the following form

$$\left.\begin{aligned}T_i &= \zeta\left(\frac{3}{\delta+2}\right)rP_{1,0}\left(\cos\theta\right),\\ T_e &= \zeta\left[r + \left(\frac{1-\delta}{2+\delta}\right)\frac{1}{r^2}\right]P_{1,0}\left(\cos\theta\right).\end{aligned}\right\} \quad (25)$$

The surface temperature is obtained as $T_s = \left[3\zeta/\left(\delta+2\right)\right]P_{1,0}\left(\cos\theta\right)$, which gives the only one non-zero coefficient of surface harmonics in $T_s$ as $T_{1,0} = 3\zeta/\left(\delta+2\right)$.

In the low $Pe_s$ limit, we use the following regular perturbation expansion for any dependent variable $\psi$ (Pak et al. 2014; Sekhar et al. 2016):

$$\psi = \psi^{(0)} + Pe_s\psi^{(Pe_s)} + Pe_s^2\psi^{(Pe_s^2)} + O\left(Pe_s^3\right), \quad (26)$$

where $\psi^{(0)}$ represents the leading-order term for $Pe_s = 0$, while $\psi^{(Pe_s)}$ and $\psi^{(Pe_s^2)}$ represent higher order contributions for $Pe_s \ll 1$. Substituting this expansion in equation (16), we obtain the governing equation for surfactant concentration as (Pak et al. 2014)

$$\text{leading-order: } \nabla^2 \Gamma^{(0)} = 0, \tag{27}$$

$$O\left(Pe_s\right): \ \nabla^2 \Gamma^{(Pe_s)} = \nabla \cdot \left( \mathbf{u}_s^{(0)} \Gamma^{(0)} \right), \tag{28}$$

$$O\left(Pe_s^2\right): \ \nabla^2 \Gamma^{(Pe_s^2)} = \nabla \cdot \left( \mathbf{u}_s^{(0)} \Gamma^{(Pe_s)} + \mathbf{u}_s^{(Pe_s)} \Gamma^{(0)} \right). \tag{29}$$

A closer look into equations (27) - (29) reveals that at each order of perturbation the surfactant concentration is independent of flow field (or surface velocity) at that order. So, we can solve for the surfactant concentration first, and then solve for flow field and droplet velocity.

To obtain the leading-order surfactant concentration (which is governed by equation (27)), we substitute $\Gamma^{(0)}\left(\theta, \varphi\right)$ in terms of surface harmonics (equation (20)) in the equation (27). The distribution of $\Gamma^{(0)}$ which satisfies the mass conservation is obtained (by using the orthogonality of the associated Legendre polynomials and equation (17)) as

$$\Gamma^{(0)} = 1, \tag{30}$$

which gives the only one non-zero coefficient of associated surface harmonics as $\Gamma_{0,0}^{(0)} = 1$. We can obtain the leading-order velocity and pressure fields by substituting equations (21) - (24) in the boundary conditions (refer to appendix A for detail expressions). To obtain the droplet velocity, we use the force-free condition as

$$\mathbf{F}_H^{(0)} = 4\pi \nabla \left( r^3 p_{-2}^{(0)} \right) = \mathbf{0}, \tag{31}$$

where the spherical solid harmonics $p_{-2}^{(0)}$ is of the form

$$p_{-2}^{(0)} = r^{-3} \left[ A_{-2,0}^{(0)} P_{2,0} \left( \cos\theta \right) + A_{-2,1}^{(0)} \cos\phi P_{2,1} \left( \cos\theta \right) + \hat{A}_{-2,1}^{(0)} \sin\phi P_{2,1} \left( \cos\theta \right) \right]. \tag{32}$$

Substituting $A_{-2,0}^{(0)}$, $A_{-2,1}^{(0)}$ and $\hat{A}_{-2,1}^{(0)}$ (expressions are given in appendix A), we obtain the components of droplet velocity in the axial (along $z$) and cross-stream (along $x$ and $y$) directions as

$$U_z^{(0)} = \left[ 1 - \left( \frac{e}{R} \right)^2 - \left( \frac{5\lambda}{3\lambda + 2} \right) \frac{2}{5R^2} \right] + \frac{2\zeta Ma_T}{(3\lambda + 2)(\delta + 2)} \Bigg\}$$

$$U_x^{(0)} = U_y^{(0)} = 0.$$

$$(33)$$

The first bracketed term in $U_z^{(0)}$ is the droplet velocity solely due to imposed Poiseuille flow, while the second term is the droplet velocity solely due to linearly varying temperature field. So, the leading-order droplet velocity is not affected by the presence of surfactants. This is due to the fact that the leading-order surfactant-induced Marangoni stress vanishes, $\nabla_s \Gamma^{(0)} = \mathbf{0}$, for a uniform distribution of surfactants. The leading-order surface velocity is obtained as

$$\mathbf{u}_s^{(0)} = \left\{ \frac{3\sin\theta Ma_T \zeta}{(\delta + 2)(3\lambda + 2)} + \frac{(\lambda + 2\sin^2\theta)e\cos\varphi}{R^2(\lambda + 1)} + \frac{5\sin^3\theta}{4R^2(\lambda + 1)} \right.$$

$$\left. - \frac{\lambda\sin\theta}{R^2(3\lambda + 2)(\lambda + 1)} \right\} \mathbf{e}_\theta - \frac{e\lambda\sin\varphi\cos\theta}{R^2(\lambda + 1)} \mathbf{e}_\varphi.$$

$$(34)$$

Now, we substitute $\Gamma^{(0)}$ and $\mathbf{u}_s^{(0)}$ in equation (28) and use the surface harmonic representation (equation (20)) for $O(Pe_s)$ surfactant concentration. The $O(Pe_s)$ surfactant concentration is obtained as

$$\Gamma^{(Pe_s)} = \Gamma_{1,0}^{(Pe_s)} P_{1,0}(\cos\theta) + \Gamma_{2,1}^{(Pe_s)} \cos\varphi P_{2,1}(\cos\theta) + \Gamma_{3,0}^{(Pe_s)} P_{3,0}(\cos\theta),$$

$$(35)$$

where the coefficients of surface harmonics are obtained as

$$\Gamma_{1,0}^{(Pe_s)} = -\frac{1}{(3\lambda + 2)} \left( \frac{3Ma_T \zeta}{\delta + 2} + \frac{2}{R^2} \right), \Gamma_{2,1}^{(Pe_s)} = -\frac{e}{3R^2(\lambda + 1)}, \Gamma_{3,0}^{(Pe_s)} = \frac{1}{6R^2(\lambda + 1)}.$$

$$(36)$$

Similar to the leading-order calculation (except there is no thermocapillary-induced Marangoni stress at the droplet interface at $O(Pe_s)$), we obtain the $O(Pe_s)$ velocity and pressure fields (refer to appendix A for detail expressions). The $O(Pe_s)$ droplet velocity is obtained from the force-free condition as

$$U_z^{(Pe_s)} = -\frac{2Ma_\Gamma}{(3\lambda + 2)^2} \left[ \frac{Ma_T \zeta}{(\delta + 2)} + \frac{2}{3R^2} \right] \Bigg\}$$

$$U_x^{(Pe_s)} = U_y^{(Pe_s)} = 0.$$

$$(37)$$

It is evident from the above expression that the surfactant-induced Marangoni stress affects the axial droplet velocity at $O(Pe_s)$.

The $O(Pe_s)$ surface velocity is obtained as

$$\mathbf{u}_s^{(Pe_s)} = Ma_\Gamma \left\{ -\frac{3\sin\theta\, Ma_T\zeta}{(\delta+2)(3\lambda+2)^2} + \frac{e\cos(2\theta)}{5R^2(\lambda+1)^2}\cos\varphi - \frac{5\sin^3\theta}{28R^2(\lambda+1)^2} \right.$$

$$\left. -\frac{4(5\lambda^2+16\lambda+8)\sin\theta}{28R^2(3\lambda+2)^2(\lambda+1)^2} \right\} \mathbf{e}_\theta - \frac{Ma_\Gamma e\sin\varphi\cos\theta}{5R^2(\lambda+1)^2}\mathbf{e}_\varphi. \tag{38}$$

Now, we substitute $\Gamma^{(0)}$, $\mathbf{u}_s^{(0)}$, $\Gamma^{(Pe_s)}$ and $\mathbf{u}_s^{(Pe_s)}$ in equation (29) and use the surface harmonic representation (equation (20)) for $O(Pe_s^2)$ surfactant concentration. The $O(Pe_s^2)$ surfactant concentration is obtained as

$$\Gamma^{(Pe_s^2)} = \left\{ \begin{array}{l} \Gamma_{1,1}^{(Pe_s^2)}P_{1,1}(\cos\theta) + \Gamma_{1,1}^{(Pe_s^2)}\cos\varphi P_{1,1}(\cos\theta) + \Gamma_{2,0}^{(Pe_s^2)}P_{2,0}(\cos\theta) + \Gamma_{2,1}^{(Pe_s^2)}\cos\varphi P_{2,1}(\cos\theta) \\ +\Gamma_{2,2}^{(Pe_s^2)}\cos(2\varphi)P_{2,2}(\cos\theta) + \Gamma_{3,0}^{(Pe_s^2)}P_{3,0}(\cos\theta) + \Gamma_{3,1}^{(Pe_s^2)}\cos\varphi P_{3,1}(\cos\theta) \end{array} \right\}, \tag{39}$$

where the coefficients of surface harmonics are mentioned in appendix B. Similar to the $O(Pe_s)$ calculation, we obtain the $O(Pe_s^2)$ velocity and pressure fields (refer to appendix A for detail expressions), and we obtain the droplet velocity using the force-free condition as

$$\left. \begin{array}{l} U_z^{(Pe_s^2)} = \dfrac{2Ma_\Gamma^2}{3(3\lambda+2)^3}\left[ \dfrac{2}{R^2} + \dfrac{3Ma_T\zeta}{(2+\delta)} \right], \\[4mm] U_x^{(Pe_s^2)} = -Ma_\Gamma e\left[ \dfrac{Ma_T\zeta(5\lambda+3)}{5R^2(3\lambda+2)^2(\lambda+1)(2+\delta)} + \dfrac{70\lambda^2+109\lambda+40}{105R^4(3\lambda+2)^2(\lambda+1)^2} \right], \\[4mm] U_y^{(Pe_s^2)} = 0. \end{array} \right\} \tag{40}$$

It is evident from the above expression that the surfactant-induced Marangoni stress not only affects the axial velocity but also induces a cross-stream migration velocity of the droplet along $x$. It is interesting to note that though the temperature gradient is acting along the axial direction, it affects the drop motion in the cross-stream direction.

### 3.3. Solution for $Pe_s \to \infty$

In the high $Pe_s$ limit, the surfactant transport equation simplifies to (Hanna & Vlahovska 2010)

$$\nabla_s \cdot \left( \mathbf{u}_s \Gamma \right) = 0, \tag{41}$$

which is still a nonlinear equation. Following Hanna & Vlahovska (2010), we consider the limiting situation of $Ma_\Gamma \gg 1$. We expand any dependent variable $\psi$ (except the surfactant concentration $\Gamma$) in the following asymptotic form (Hanna & Vlahovska 2010; Schwalbe et al. 2011):

$$\psi = \psi^{(0)} + Ma_\Gamma^{-1} \psi^{\left( Ma_\Gamma^{-1} \right)} + O\left( Ma_\Gamma^{-2} \right), \tag{42}$$

where $\psi^{(0)}$ represents the leading-order term for $Ma_\Gamma \to \infty$, and $\psi^{\left( Ma_\Gamma^{-1} \right)}$ represents the higher order correction for $Ma_\Gamma^{-1} \ll 1$. The surfactant concentration is expanded in the following form (Hanna & Vlahovska 2010; Schwalbe et al. 2011)

$$\Gamma = 1 + Ma_\Gamma^{-1} \Gamma^{(0)} + Ma_\Gamma^{-2} \Gamma^{\left( Ma_\Gamma^{-1} \right)} + O\left( Ma_\Gamma^{-3} \right). \tag{43}$$

Substituting equations (42) and (43) in equation (41), we obtain the governing equation for surfactant concentration at different orders of perturbation as

$$\text{leading-order: } \nabla \cdot \mathbf{u}_s^{(0)} = 0, \tag{44}$$

$$O\left( Ma_\Gamma^{-1} \right): \ \ \nabla \cdot \left( \mathbf{u}_s^{(0)} \Gamma^{(0)} + \mathbf{u}_s^{\left( Ma_\Gamma^{-1} \right)} \right) = 0. \tag{45}$$

The tangential stress balance at droplet surface $\left( r = 1 \right)$ can be obtained as

$$\text{leading-order: } \left( \boldsymbol{\tau}_e^{(0)} \cdot \mathbf{e}_r - \boldsymbol{\tau}_i^{(0)} \cdot \mathbf{e}_r \right) \cdot \left( \mathbf{I} - \mathbf{e}_r \mathbf{e}_r \right) = \nabla_s \Gamma^{(0)} + Ma_T \nabla_s T_s \tag{46}$$

$$O\left( Ma_\Gamma^{-1} \right): \ \ \left( \boldsymbol{\tau}_e^{\left( Ma_\Gamma^{-1} \right)} \cdot \mathbf{e}_r - \boldsymbol{\tau}_i^{\left( Ma_\Gamma^{-1} \right)} \cdot \mathbf{e}_r \right) \cdot \left( \mathbf{I} - \mathbf{e}_r \mathbf{e}_r \right) = \nabla_s \Gamma^{\left( Ma_\Gamma^{-1} \right)}. \tag{47}$$

A closer look into equations (44) - (47) reveals that the surfactant concentration and flow field have to be obtained simultaneously. Similar to low $Pe_s$ limit, the temperature field is independent of flow field and surfactant concentration. So, the solution for temperature field will be same as given in equation (25). Now, we substitute the velocity and pressure fields given in equations (21) - (24) in the boundary conditions (refer to appendix C for detail expressions for different spherical harmonics present in the flow field). The leading-order surfactant concentration and surface velocity are obtained as

$$
\left.\begin{array}{l}
\Gamma^{(0)} = -\dfrac{\cos\theta}{12(\delta+2)}\left[36 Ma_T\zeta + \dfrac{(\delta+2)}{R^2}\left(24 + 60e\cos\varphi\sin\theta - 35e^2\cos^2\theta + 21e^2\right)\right], \\[3mm]
\mathbf{u}_s^{(0)} = \left(\dfrac{e\cos\varphi}{R^2}\right)\mathbf{e}_\theta - \dfrac{e\sin\varphi\cos\theta}{R^2}\mathbf{e}_\varphi.
\end{array}\right\} \tag{48}
$$

Force-free conditions gives the leading-order droplet velocity as

$$
U_z^{(0)} = 1 - \frac{2}{3R^2} - \left(\frac{e}{R}\right)^2 , \ \ U_x^{(0)} = U_y^{(0)} = 0. \tag{49}
$$

It is evident from the above expression that the surfactant-induced Marangoni stress leads to motion of a droplet at a speed of solid particle. Thermocapillarity has no effect on the droplet speed at the leading-order approximation.

The $O\left(Ma_\Gamma^{-1}\right)$ surfactant concentration is obtained as

$$
\Gamma^{(Ma_\Gamma^{-1})} = \left[\begin{array}{l}
\Gamma_{1,1}^{(Ma_\Gamma^{-1})}\cos\varphi P_{1,1}\left(\cos\theta\right) + \Gamma_{2,0}^{(Ma_\Gamma^{-1})}P_{2,0}\left(\cos\theta\right) \\[2mm]
+\Gamma_{2,2}^{(Ma_\Gamma^{-1})}\cos 2\varphi P_{2,2}\left(\cos\theta\right) + \Gamma_{3,1}^{(Ma_\Gamma^{-1})}\cos\varphi P_{3,1}\left(\cos\theta\right)
\end{array}\right], \tag{50}
$$

where the coefficients of surface harmonics are of the form

$$
\left.\begin{array}{l}
\Gamma_{1,1}^{(Ma_\Gamma^{-1})} = \dfrac{3}{2}\dfrac{e\left(4 + 3Ma_T\zeta R^2 + 2\delta\right)(\lambda+1)}{R^4\left(\delta+2\right)}, \ \ \Gamma_{0,2}^{(Ma_\Gamma^{-1})} = -\dfrac{25}{6}\dfrac{e^2\left(\lambda+1\right)}{R^4}, \\[3mm]
\Gamma_{2,2}^{(Ma_\Gamma^{-1})} = \dfrac{25}{36}\dfrac{e^2\left(\lambda+1\right)}{R^4}, \ \ \Gamma_{1,3}^{(Ma_\Gamma^{-1})} = -\dfrac{49}{72}\dfrac{e^3\left(\lambda+1\right)}{R^4}.
\end{array}\right\} \tag{51}
$$

Force-free conditions gives the $O\left(Ma_\Gamma^{-1}\right)$ droplet velocity as

$$
U_z^{(Ma_\Gamma^{-1})} = U_y^{(Ma_\Gamma^{-1})} = 0, \ \ U_x^{(Ma_\Gamma^{-1})} = -\frac{e\left(4 + 3Ma_T\zeta R^2 + 2\delta\right)}{3R^4\left(\delta+2\right)}. \tag{52}
$$

Equation (52) shows that the surfactant-induced Marangoni stress induces a non-zero cross-stream migration of the droplet along $x$. Importantly, thermocapillarity affects the cross-stream velocity of the droplet at this order of approximation.

## 4. Results and discussion

### 4.1. *Effect of Marangoni stress in the low $Pe_s$ limit*

The main result of the present analysis is the velocity of a force-free surfactant-laden droplet in the combined presence of Poiseuille flow and linearly varying temperature field. We have obtained the droplet velocity in low $Pe_s$ limit as

$$
\mathbf{U} = \left[ \left\{ 1 - \left( \frac{e}{R} \right)^2 - \frac{2\lambda}{(3\lambda+2)R^2} + \frac{2\zeta Ma_T}{(3\lambda+2)(\delta+2)} \right\} - Pe_s \left\{ \frac{2Ma_\Gamma}{(3\lambda+2)^2} \left( \frac{Ma_T\zeta}{\delta+2} + \frac{2}{3R^2} \right) \right\} \right.
$$

$$
\left. + Pe_s^2 \left\{ \frac{2Ma_\Gamma^2}{3(3\lambda+2)^3} \left( \frac{2}{R^2} + \frac{3Ma_T\zeta}{2+\delta} \right) \right\} \right] \mathbf{e}_z - \left[ Pe_s^2 Ma_\Gamma e \left\{ \frac{Ma_T\zeta(5\lambda+3)}{5R^2(3\lambda+2)^2(\lambda+1)(2+\delta)} \right. \right.  \quad (53)
$$

$$
\left. \left. + \frac{70\lambda^2 + 109\lambda + 40}{105R^4(3\lambda+2)^2(\lambda+1)^2} \right\} \right] \mathbf{e}_x + O\left( Pe_s^3 \right).
$$

Substituting $Ma_T = 0$ in equation (53), we recover the velocity of a surfactant-laden droplet in isothermal Poiseuille flow in the low $Pe_s$ limit which was previously obtained by Pak et al. (2014). Substituting $e = 0$ in equation (53), we obtain the axial velocity of a surfactant-laden droplet in non-isothermal Poiseuille flow which was previously obtained by Das et al. (2016). A closer look into equation (53) reveals that the thermocapillary-induced Marangoni stress in the presence of surfactant-induced Marangoni stress alters both axial and cross-stream velocities of the droplet.

Now, we investigate the effects of Marangoni stresses on the axial velocity of the droplet. Towards this, we first plot figure 2 which shows the variation of droplet velocity in the axial direction $(U_z)$ with the viscosity ratio $(\lambda)$ for the particular case in which the temperature increases in the direction of Poiseuille flow $(\text{i.e.} \zeta = 1)$. $Ma_T = 0$ represents the special case of surfactant-laden droplet in isothermal flow which. When $Ma_T > 0$ figure 2 depicts a significant increase in magnitude of axial droplet velocity. This is due to the fact that for $\zeta = 1$ the thermocapillary-induced Marangoni stress drives the droplet in the direction of Poiseuille flow. The effect of Marangoni stress diminishes in the limit $\lambda \to \infty$ which is due to the fact that the droplet surface becomes more rigid and the interfacial tension plays no role in governing the hydrodynamics.

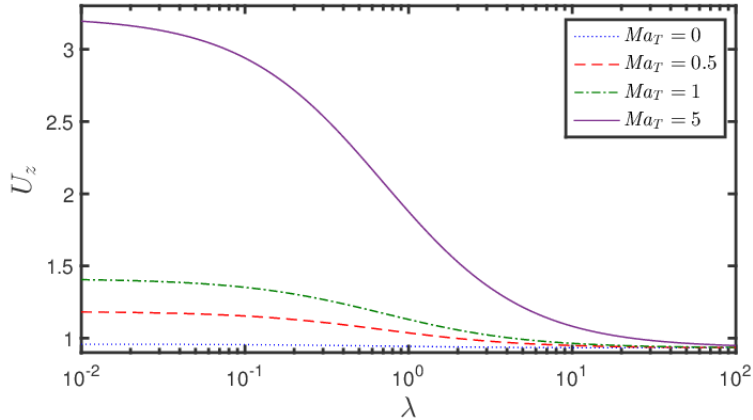

FIGURE 2. The variation of axial droplet velocity $\left(U_z\right)$ with the viscosity ratio $\left(\lambda\right)$ for different values of $Ma_T$. Other parameter have the following values: $\zeta = 1$, $Ma_\Gamma = 1$, $Pe_s = 0.1$, $\delta = 0.1$, $e = 1$ and $R = 5$.

We next consider the special case of $\zeta = -1$ which signifies the decreasing temperature field in the imposed Poiseuille flow direction. Figure 3 shows the variation of droplet velocity in the axial direction $\left(U_z\right)$ with the viscosity ratio $\left(\lambda\right)$ for different values of $Ma_T$. In this case the thermocapillary-induced Marangoni stress acts to drive the droplet in the direction opposite to the Poiseuille flow. For smaller values of $Ma_T$ (e.g. $Ma_T = 0.5$), the effect of imposed Poiseuille flow dominates and the droplet moves in the direction of Poiseuille flow $\left(\text{i.e.} U_z > 0\right)$. For larger values of $Ma_T$ (e.g. $Ma_T = 5$) the droplet moves in the direction of Poiseuille flow $\left(\text{i.e.} U_z > 0\right)$ for larger viscosity ratio, while low viscosity droplet moves opposite to the Poiseuille flow $\left(\text{i.e.} U_z < 0\right)$. For a given viscosity ratio $\left(\lambda\right)$, we obtain a critical value of $Ma_T$ for which $U_z = 0$ in the following form

$$Ma_{T,z}^* = \frac{\left[3\left(3\lambda+2\right)^2\left\{\left(R^2-e^2\right)\left(3\lambda+2\right)-2\lambda\right\}-4Ma_\Gamma\left(3\lambda+2\right)Pe_s+4Ma_\Gamma^2 Pe_s^2\right]\left(\delta+2\right)}{6R^2\left\{\left(3\lambda+2\right)^2-Ma_\Gamma\left(3\lambda+2\right)Pe_s+Ma_\Gamma^2 Pe_s^2\right\}}. \quad (54)$$

Important to note that for $Ma_T < Ma_{T,z}^*$ the magnitude of axial droplet velocity decreases with increase in $Ma_T$, while the magnitude of axial droplet velocity increases with increase in $Ma_T$ for $Ma_T > Ma_{T,z}^*$.

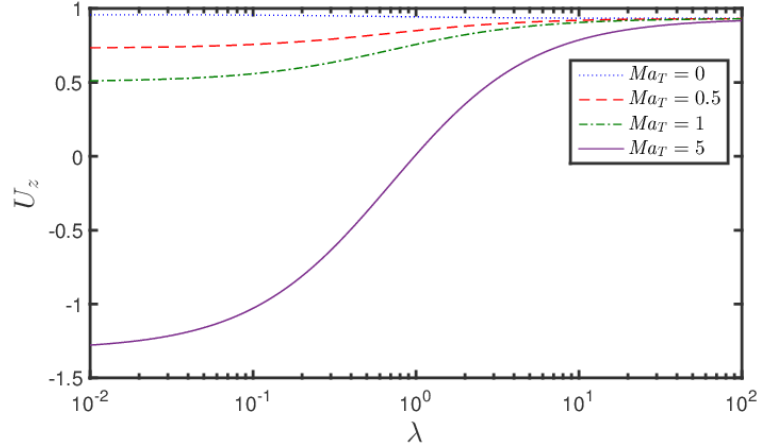

FIGURE 3. The variation of axial droplet velocity $\left(U_z\right)$ with the viscosity ratio $\left(\lambda\right)$ for the special case of $\zeta = -1$. Other parameters have the following values: $Ma_\Gamma = 1$, $Pe_s = 0.1$, $\delta = 0.1$, $e = 1$ and $R = 5$.

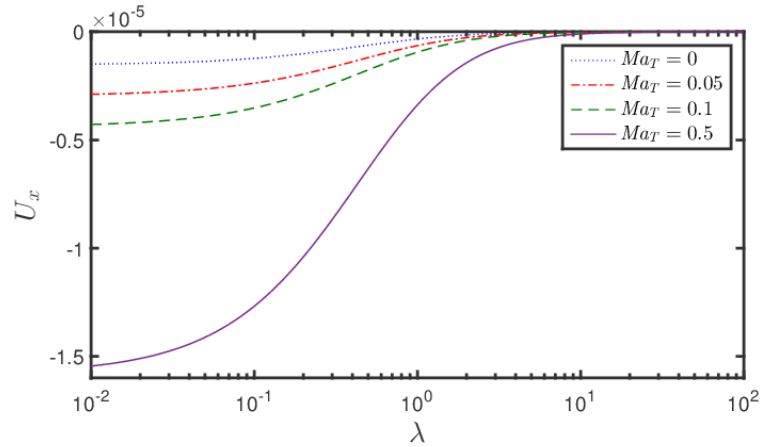

FIGURE 4. The variation of cross-stream droplet velocity $\left(U_x\right)$ with the viscosity ratio $\left(\lambda\right)$ for different values of $Ma_T$. Other parameters have the following values: $\zeta = 1$, $Ma_\Gamma = 1$, $Pe_s = 0.1$, $\delta = 0.1$, $e = 1$ and $R = 5$.

Now, we investigate the effects of Marangoni stresses on the cross-stream velocity of the droplet. Towards this, we first plot figure 4 which shows the variation of droplet velocity in the cross-stream direction $\left(U_x\right)$ with the viscosity ratio $\left(\lambda\right)$ for the case in which the temperature increases in the direction of Poiseuille flow $\left(\text{i.e.}\, \zeta = 1\right)$. For isothermal Poiseuille flow $\left(\text{i.e.}\, Ma_T = 0\right)$, Pak et al. (2014) obtained cross-stream motion of a surfactant-laden droplet. They have obtained that an eccentrically positioned droplet always moves towards the channel centerline (represented by negative values of $U_x$ in figure 4). Figure 4 depicts that in the presence of temperature variation (i.e. $Ma_T > 0$ with increasing temperature in the direction of imposed Poiseuille flow), there is a significant increases in the magnitude of cross-stream velocity $\left(\text{i.e.} \left|U_x\right|\right)$. So the thermocapillary-induced Marangoni stress aids the surfactant-induced cross-stream migration of the droplet for $\zeta = 1$. The magnitude of cross-stream velocity increases with increase in $Ma_T$ for low viscosity droplets. Similar to axial velocity, the effect of $Ma_T$ (i.e. thermocapillary-induced Marangoni stress) on the cross-stream velocity of droplet diminishes for larger viscosity ratio $\left(\text{i.e.} \lambda \to \infty\right)$.

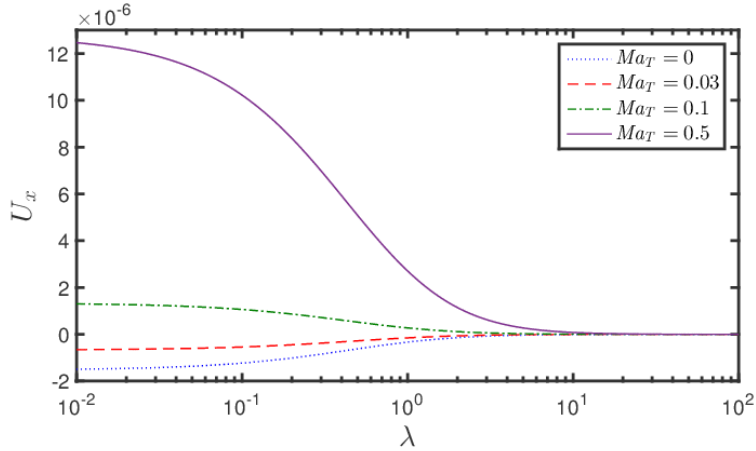

FIGURE 5. The variation of cross-stream droplet velocity $\left(U_x\right)$ with the viscosity ratio $\left(\lambda\right)$ for different values of $Ma_T$. Other parameters have the following values: $\zeta = -1$, $Ma_\Gamma = 1$, $Pe_s = 0.1$, $\delta = 0.1$, $e = 1$ and $R = 5$.

We next consider the special case of $\zeta = -1$ which signifies the decreasing temperature field in the imposed Poiseuille flow direction. Figure 5 shows the variation of droplet velocity in the cross-stream direction $\left(U_x\right)$ with the viscosity ratio $\left(\lambda\right)$ for different values of $Ma_T$. In this case the thermocapillary-induced Marangoni stress not only alters the magnitude of cross-stream migration velocity but also alters the direction of cross-stream migration velocity. For smaller values of $Ma_T$ (e.g. $Ma_T = 0.03$) the droplet moves towards the channel centerline $\left(\text{i.e.} U_x < 0\right)$. However, for larger values of $Ma_T$ (e.g. $Ma_T = 0.5$) the droplet moves away from the channel centerline $\left(\text{i.e.} U_x > 0\right)$. One can find a critical value of $Ma_T$ for which $U_x = 0$ in the following form

$$Ma_{T,x}^* = \frac{\left(\delta + 2\right)\left(70\lambda^2 + 109\lambda + 40\right)}{21\left(5\lambda + 3\right)\left(\lambda + 1\right) R^2}.$$

(55)

Important to note that for $Ma_T < Ma_{T,x}^*$ the magnitude of cross-stream velocity decreases with increase in $Ma_T$, while the magnitude of cross-stream velocity increases with increase in $Ma_T$ for $Ma_T > Ma_{T,x}^*$.

The physical reason of the change in magnitude and direction of axial and cross-stream velocities of a surfactant-laden droplet in non-isothermal Poiseuille flow can be understood by looking into the distribution of surfactant concentration and interfacial tension on the droplet surface. Towards this, we first look into the case of droplet located at flow centerline (i.e. $e = 0$). Figure 6(a) and 6(b) depict the distribution of $\Gamma(\theta, \varphi)$ and $\sigma(\theta, \varphi)$ for the particular case of $Ma_T = 0$ (i.e. isothermal Poiseuille flow). Imposed Poiseuille flow alters the surfactant distribution from its equilibrium value $\Gamma_{eq} = 1$. Surfactant concentration is less near the east pole as compared with the west pole as depicted in figure 6(a). So, the altered surfactant distribution is asymmetric about the transverse plane. This distribution is due to the imposed Poiseuille flow which drives the surfactants from the east pole to the west pole. The asymmetry in $\Gamma(\theta, \varphi)$ leads to asymmetric distribution of $\sigma(\theta, \varphi)$ about the transverse plane which is depicted in figure 6(b). Larger interfacial tension near the east pole as compared with the west pole drives a surfactant-induced Marangoni flow from the west to east pole, which leads to the generation of a hydrodynamic forces in the axial direction opposite to the Poiseuille flow. This reduces the droplet speed in the axial direction as compared with the speed of a surfactant-free droplet. The distribution of $\Gamma(\theta, \varphi)$ and $\sigma(\theta, \varphi)$ remain symmetric about the axial plane which results in no motion of the droplet in the cross-stream direction. The distributions of $\Gamma(\theta, \varphi)$ and $\sigma(\theta, \varphi)$ in non-isothermal Poiseuille flow (i.e. $Ma_T > 0$) for $\zeta = 1$ (i.e. increasing temperature in the direction of Poiseuille flow) are shown in figure 6(c) and 6(d), respectively. For $\zeta = 1$ the thermocapillary-induced Marangoni stress induces a flow at the droplet surface which runs from the east to west pole. This flow drives surfactants away from the east pole towards the west pole which leads to an increases in the asymmetry in $\Gamma(\theta, \varphi)$. Comparison of figure 6(c) with 6(a) reflects the increase in asymmetry in terms of increase in $(\Gamma_{max} - \Gamma_{min})$. This surfactant distribution combined with the nonuniform temperature distribution leads to larger $\sigma(\theta, \varphi)$ near the west pole as compared with the east pole (refer to figure 6(d)) which is in sharp contrast to the case of $Ma_T = 0$ shown in figure 6(b). This is due to the fact that the east pole encounters relatively hot fluid (higher temperature means lower interfacial tension). This distribution of $\sigma(\theta, \varphi)$ drives a Marangoni flow from the east to west pole which leads to motion of droplet in the direction of Poiseuille flow with increased velocity as compared with $Ma_T = 0$.

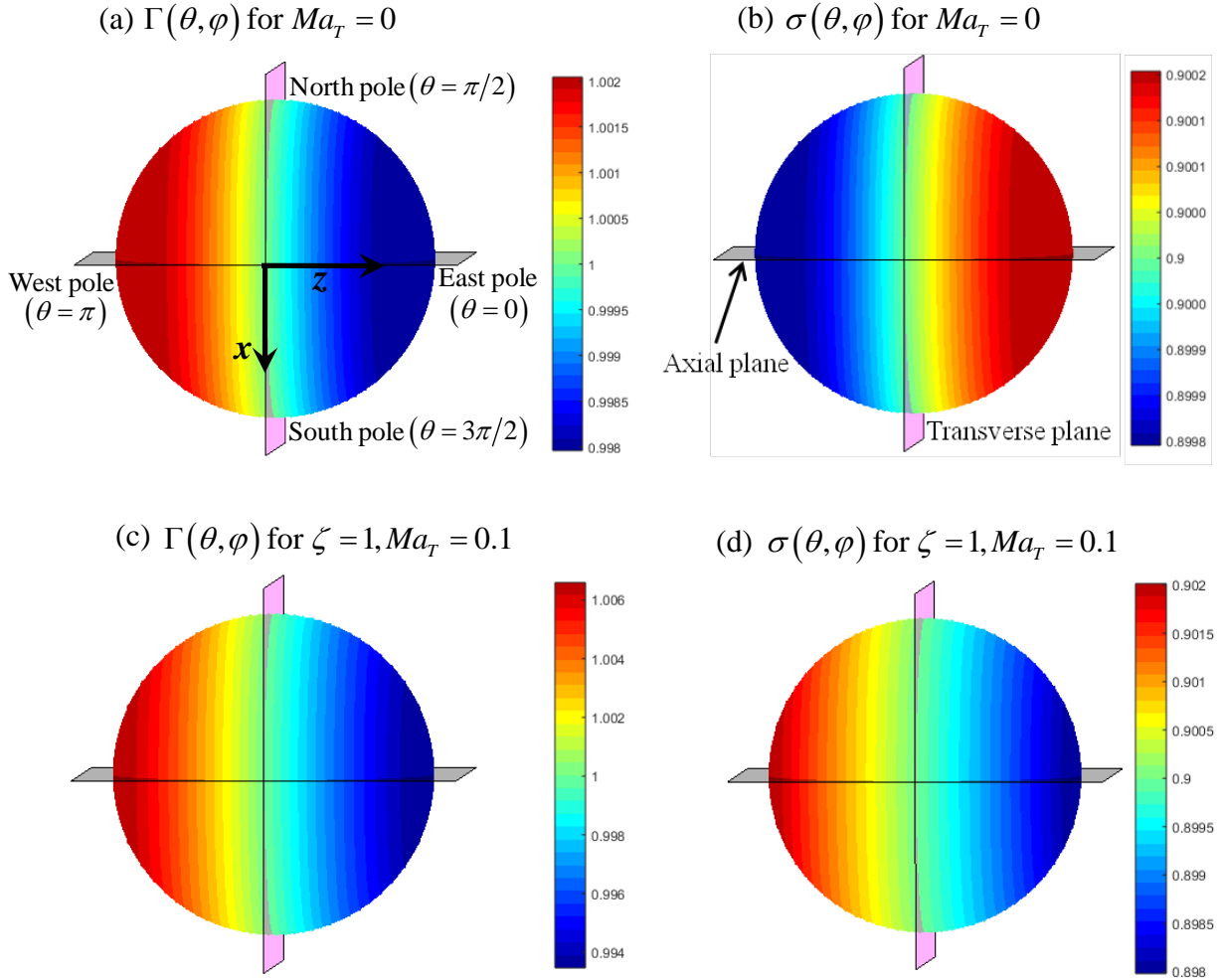

FIGURE 6. Surface plot showing the variation of surfactant concentration for (a) $Ma_T = 0$, (c) $\zeta = 1$ and $Ma_T = 0.1$. Surface plot showing the variation of interfacial tension for (b) $Ma_T = 0$, (d) $\zeta = 1$ and $Ma_T = 0.1$. Other parameters have the following values: $Ma_\Gamma = 5$, $Pe_s = 0.1$, $\delta = 0.1$, $\lambda = 0.1$, $e = 0$ and $R = 5$.

Now, we consider the case of for $\zeta = -1$ (i.e. decreasing temperature in the direction of Poiseuille flow) and $e = 0$. For $\zeta = -1$ the thermocapillary-induced Marangoni stress induces a flow at the droplet surface which runs from the west to east pole. This flow drives surfactants from the west to the east pole, while the imposed Poiseuille flow drives the surfactants in the opposite direction (i.e. from the east to west pole). The final surfactant distribution is decided by the net surface velocity. For $Ma_T = 0.5$, the distribution of $\Gamma(\theta, \varphi)$ and $\sigma(\theta, \varphi)$ are shown in figure 7(a) and 7(b), respectively. Figure 7(a) depicts that the combined effect of thermocapillary-induced Marangoni stress and imposed Poiseuille flow leads to larger surfactant concentration near the east pole as compared with the west pole. This surfactant distribution combined with the nonuniform temperature distribution leads to larger $\sigma(\theta, \varphi)$ near the east pole as compared with the west pole as depicted in figure 7(b). This Marangoni stress induces a hydrodynamic force in the direction opposite to the Poiseuille flow. For $Ma_T = 0.5$ the Poiseuille flow dominates and the droplet moves in the direction of Poiseuille flow but with decreased axial velocity of the droplet. Now, we consider $Ma_T = 5$ and plot $\Gamma(\theta, \varphi)$ and $\sigma(\theta, \varphi)$ in figure 7(c) and 7(d), respectively. With larger $Ma_T$, the strong effect of thermocapillary-induced Marangoni stress leads to significant increase (and decrease) in surfactant concentration near east (and west) pole. Comparison between figure 7(a) and 7(c) reveals that $(\Gamma_{max} - \Gamma_{min})$ is much larger for $Ma_T = 5$. Similar increase in asymmetry is also reflected in figure 7(d) for $\sigma(\theta,\varphi)$. In this case the Marangoni stress dominates over the Poiseuille flow and the droplet moves in the direction opposite to the Poiseuille flow.

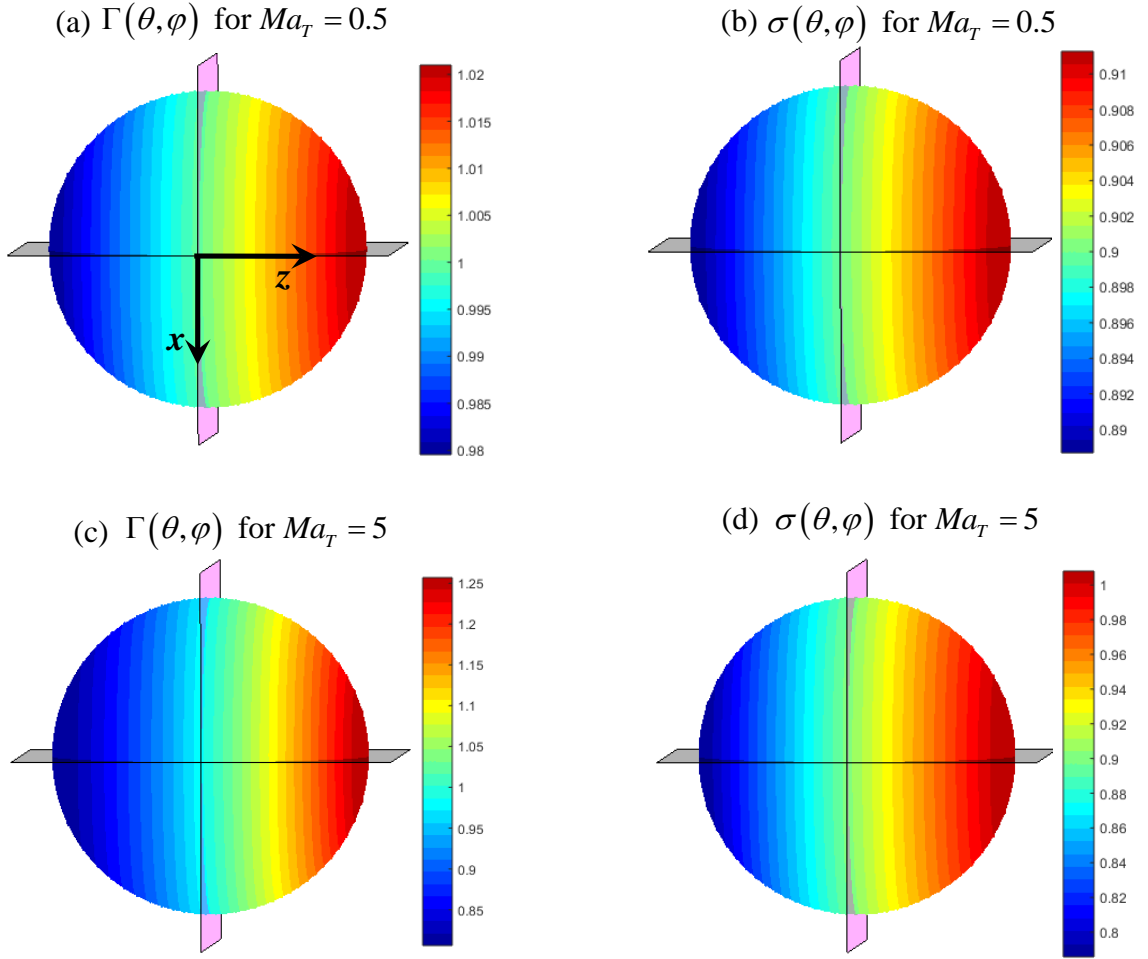

FIGURE 7. Surface plot showing the variation of surfactant concentration for (a) $Ma_T = 0.5$, (c) $Ma_T = 5$. Surface plot showing the variation of interfacial tension for (b) $Ma_T = 0.5$, (d) $Ma_T = 5$. Other parameters have the following values: $\zeta = -1$, $Ma_\Gamma = 5$, $Pe_s = 0.1$, $\delta = 0.1$, $\lambda = 0.1$, $e = 0$ and $R = 5$.

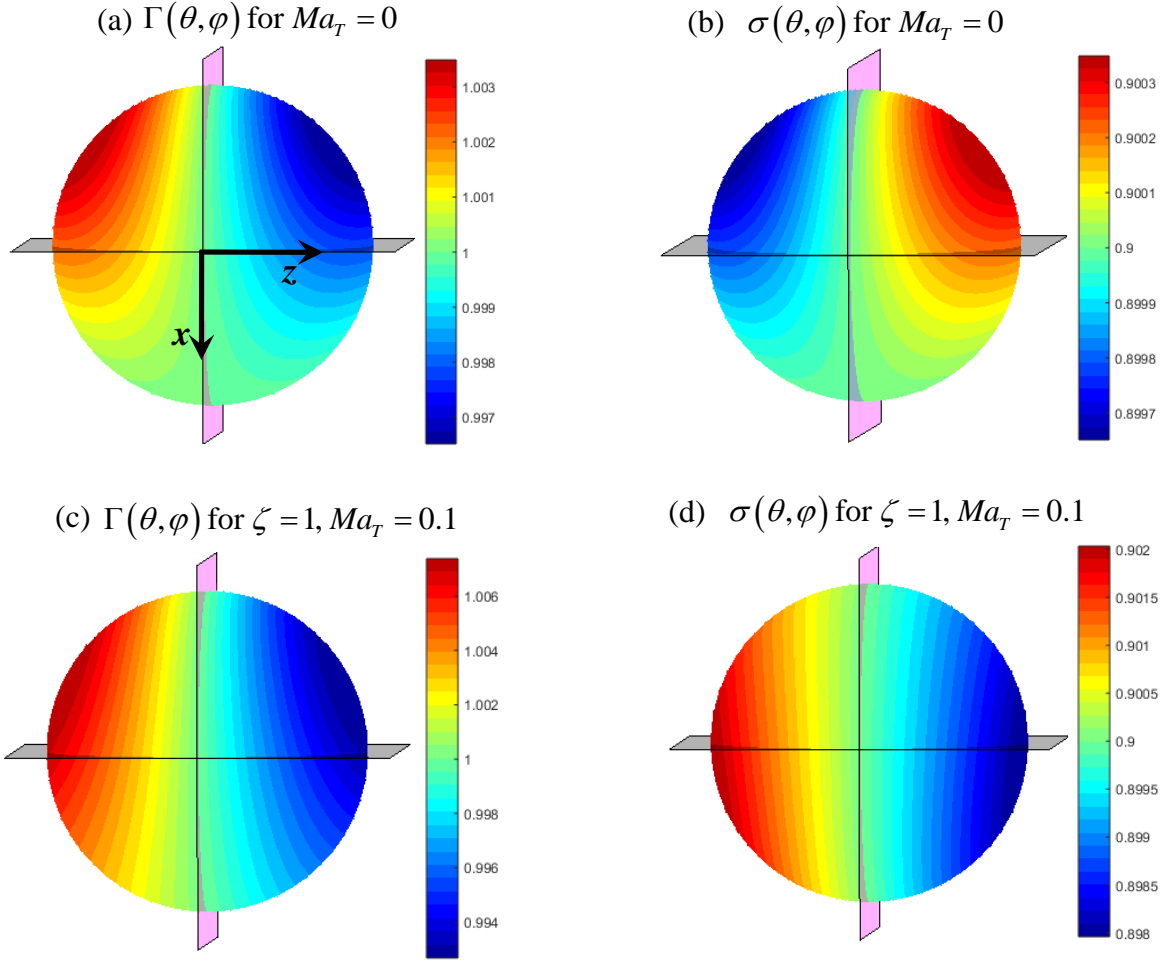

(a) $\Gamma(\theta,\varphi)$ for $Ma_T = 0$

(b) $\sigma(\theta,\varphi)$ for $Ma_T = 0$

(c) $\Gamma(\theta,\varphi)$ for $\zeta = 1$, $Ma_T = 0.1$

(d) $\sigma(\theta,\varphi)$ for $\zeta = 1$, $Ma_T = 0.1$

FIGURE 8. Surface plot showing the variation of surfactant concentration for (a) $Ma_T = 0$, (c) $\zeta = 1$ and $Ma_T = 0.1$. Surface plot showing the variation of interfacial tension for (b) $Ma_T = 0$, (d) $\zeta = 1$ and $Ma_T = 0.1$. The other parameters are $Ma_\Gamma = 5$, $Pe_s = 0.1$, $\delta = 0.1$, $\lambda = 0.1$, $e = 1$ and $R = 5$.

Interesting things happen for $e = 1$ (i.e. eccentrically located droplet). The distributions of $\Gamma(\theta,\varphi)$ and $\sigma(\theta,\varphi)$ in isothermal Poiseuille flow (i.e. $Ma_T = 0$) for $e = 1$ and $\zeta = 1$ are shown in figure 8(a) and 8(b), respectively. Figure 8(a) depicts that the surfactant concentration is maximum near northwest pole and minimum near the northeast pole. So, now the distribution of $\Gamma(\theta,\varphi)$ is asymmetric about both the transverse and axial planes. The asymmetry about the axial plane is due to the asymmetric velocity field about the axial plane encountered by the eccentrically located droplet. An eccentrically located droplet encounters larger velocity in the north pole (as it is closer to channel centerline) and smaller velocity in the south pole (as it is away from the channel centerline) as depicted in figure 1. Asymmetry in $\Gamma(\theta,\varphi)$ creates asymmetry in $\sigma(\theta,\varphi)$ (refer to figure 8(b)) which leads to the generation of hydrodynamic forces in both axial and transverse directions which causes the cross-stream migration of eccentrically located droplet. The distributions of $\Gamma(\theta,\varphi)$ and $\sigma(\theta,\varphi)$ in non-isothermal Poiseuille flow (i.e. $Ma_T > 0$) for $\zeta = 1$ (i.e. increasing temperature in the direction of Poiseuille flow) and $e = 1$ are shown in figure 8(c) and 8(d), respectively. For $\zeta = 1$ the thermocapillary-induced Marangoni flow drives surfactants from the east pole to west pole and increase the asymmetry in $\Gamma(\theta,\varphi)$ which leads to maximum surfactant concentration near the northwest pole and minimum surfactant concentration near the northeast pole. Comparison of figure 8(c) with 8(a) reflects the increase in asymmetry in terms of increase in $(\Gamma_{\max} - \Gamma_{\min})$. This surfactant distribution combined with the nonuniform temperature distribution leads to larger $\sigma(\theta,\varphi)$ near the west pole as compared with the east pole (refer to figure 8(d)). Similar to $\Gamma(\theta,\varphi)$, increase in asymmetry about the axial plane is also present in $\sigma(\theta,\varphi)$ (i.e. $(\sigma_{\max} - \sigma_{\min})$) which further leads to increase in magnitude of cross-stream velocity of the droplet.

<div align="center">

(a) $\Gamma(\theta,\varphi)$ for $Ma_T = 0.01$          (b) $\Gamma(\theta,\varphi)$ for $Ma_T = 0.1$

</div>

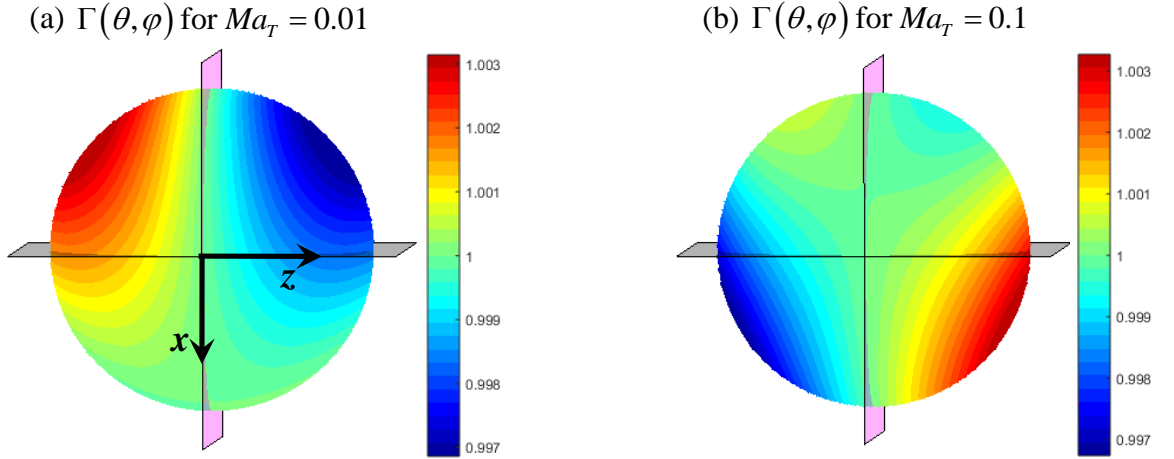

FIGURE 9. Surface plot showing the variation of surfactant concentration for (a) $Ma_T = 0.01$ and (b) $Ma_T = 0.1$. Other parameters have the following values: $\zeta = -1$, $Ma_\Gamma = 5$, $Pe_s = 0.1$, $\delta = 0.1$, $\lambda = 0.1$, $e = 1$ and $R = 5$.

The distributions of $\Gamma(\theta,\varphi)$ in non-isothermal Poiseuille flow (i.e. $Ma_T > 0$) for $\zeta = -1$ (i.e. decreasing temperature in the direction of Poiseuille flow) and $e = 1$ is shown in figure 9. For $\zeta = -1$ the thermocapillary-induced Marangoni flow drives surfactants from the west to east pole, while the imposed Poiseuille flow drives the surfactant in the opposite directions (i.e. from the east to west pole). The final surfactant distribution is decided by the net surface velocity. For $Ma_T < Ma^*_{T,x}$, the Poiseuille flow dominates over the thermocapillary-induced Marangoni flow and leads maximum surfactant concentration near northwest pole and minimum surfactant concentration near northeast pole (refer to figure 9(a)). This distribution is similar to the case of $\zeta = 1$ (refer to figure 8(c)) which leads to cross-stream motion of droplet towards the channel centerline. However, when thermocapillary-induced Marangoni stress dominates over the Poiseuille flow (for $Ma_T > Ma^*_{T,x}$), the surfactant concentration is maximum near southeast pole and minimum near southwest pole (refer to figure 9(b)). This is completely opposite to that of $Ma_T < Ma^*_{T,x}$ (shown in figure 9(a)). This asymmetric distribution of $\Gamma(\theta, \varphi)$ leads to cross-stream migration of the droplet away from the channel centerline.

### 4.2. Effect of Marangoni stress in the high $Pe_s$ limit

We have obtained the droplet velocity in the high $Pe_s$ limit as

$$\mathbf{U} = \left[ 1 - \frac{2}{3R^2} - \left( \frac{e}{R} \right)^2 \right] \mathbf{e}_z - Ma_\Gamma^{-1} \left[ \frac{e\left( 4 + 3Ma_T \zeta R^2 + 2\delta \right)}{3R^4\left( \delta + 2 \right)} \right] \mathbf{e}_x + O\left( Ma_\Gamma^{-2} \right). \qquad (56)$$

Substituting $Ma_T = 0$ in equation (56), we recover the velocity of a surfactant-laden droplet in isothermal Poiseuille flow in the large $Pe_s$ limit which was previously obtained by Hanna & Vlahovska (2010). A closer look into equation (56) reveals that the thermocapillary-induced Marangoni stress only affects the cross-stream velocity of the droplet.

Now, we investigate the effects of Marangoni stresses on the cross-stream velocity of the droplet. Towards this, we first plot figure 10(a) which shows the variation of droplet velocity in the cross-stream direction $(U_x)$ with the conductivity ratio $(\delta)$ for the particular case in which the temperature increases in the direction of Poiseuille flow (i.e. $\zeta = 1$). Figure 10(a) depicts that in the presence of temperature variation (i.e. $Ma_T > 0$), there is a significant increases in the magnitude of cross-stream velocity $(\text{i.e.} |U_x|)$. So the thermocapillary-induced Marangoni stress aids the surfactant-induced cross-stream migration of the droplet for $\zeta = 1$. The magnitude of cross-stream velocity increases with increase in $Ma_T$ for low conductivity droplets. With increase in thermal conductivity of the droplet, the cross-stream velocity reduces. Interesting thing happens for $\zeta = -1$ (i.e. decreasing temperature in the direction of Poiseuille flow) which is depicted in figure 10(b). Similar to low $Pe_s$ limit, the thermocapillary-induced Marangoni stress not only alters the magnitude but also alters the direction of cross-stream migration for $\zeta = -1$. For smaller values of $Ma_T$ (e.g. $Ma_T = 0.01$) the droplet moves towards the channel centerline $(\text{i.e.} U_x < 0)$. However, for larger values of $Ma_T$ (e.g. $Ma_T = 1$) the droplet moves away from the channel centerline $(\text{i.e.}\, U_x > 0)$. One can find a critical value of $Ma_T$ for which $U_x = 0$ in the following form

$$Ma_{T,x}^* = \frac{2}{3R^2}(\delta + 2). \qquad (57)$$

Important to note that for $Ma_T < Ma_{T,x}^*$ the magnitude of cross-stream velocity decreases with increase in $Ma_T$, while the magnitude of cross-stream velocity increases with increase in $Ma_T$ for $Ma_T > Ma_{T,x}^*$. Another important thing to note from figure 10 is that the magnitude of cross-stream velocity for high $Pe_s$ limit is much larger as compared with the magnitude of cross-stream velocity for low $Pe_s$ limit (refer to figures 4 and 5).

(a) $\zeta = 1$

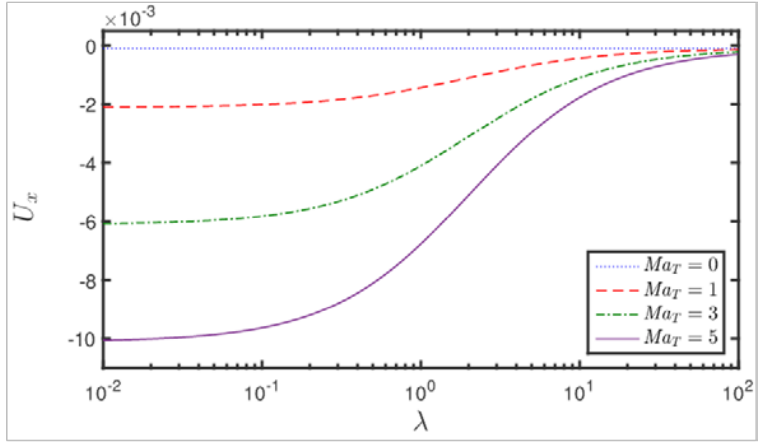

(b) $\zeta = -1$

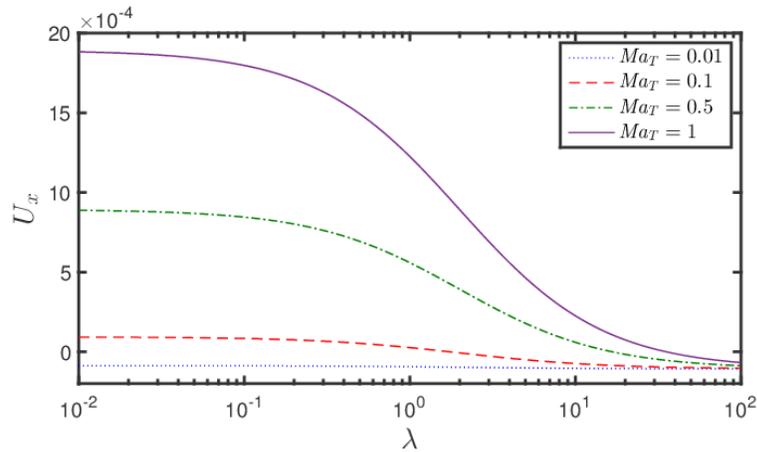

FIGURE 10. Variation of cross-stream velocity $(U_x)$ with the conductivity ratio $(\delta)$ for (a) $\zeta = 1$ and (b) $\zeta = -1$. Other parameters have the following values: $Ma_\Gamma = 10$, $e = 1$, and $R = 5$.

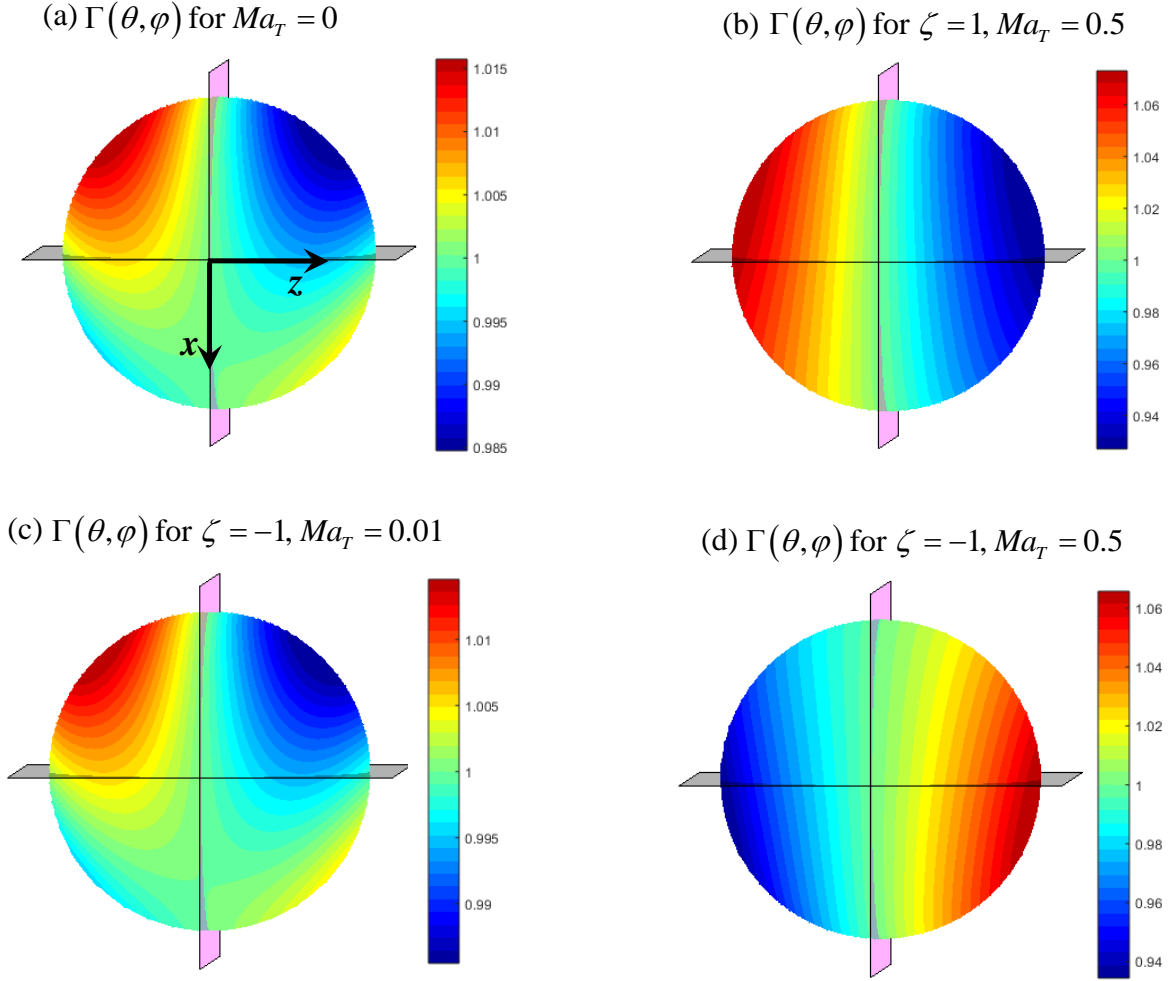

(a) $\Gamma(\theta,\varphi)$ for $Ma_T = 0$

(b) $\Gamma(\theta,\varphi)$ for $\zeta = 1, \, Ma_T = 0.5$

(c) $\Gamma(\theta,\varphi)$ for $\zeta = -1, \, Ma_T = 0.01$

(d) $\Gamma(\theta,\varphi)$ for $\zeta = -1, \, Ma_T = 0.5$

FIGURE 11. Surface plot showing the surfactant distribution for (a) $Ma_T = 0$, (b) $\zeta = 1$ and $Ma_T = 0.5$, (c) $\zeta = -1$ and $Ma_T = 0.01$, (d) $\zeta = -1$ and $Ma_T = 0.5$. Other parameters have the following values: $Ma_\Gamma = 10$, $\lambda = 0.1$, $\delta = 0.1$, $e = 1$, and $R = 5$.

Similar to the low $Pe_s$ limit, to investigate the alteration in magnitude and direction of droplet velocity in axial and cross-stream directions, we plot distribution of $\Gamma(\theta,\varphi)$ on the surface of an eccentrically located droplet for isothermal Poiseuille flow (with $Ma_T = 0$) in figure 11(a) and for non-isothermal Poiseuille flow (with $Ma_T = 0.5$ and $\zeta = 1$) in figure 11(b). Figure 11(a) depict that $\Gamma(\theta,\varphi)$ is asymmetric about both axial and transverse planes which results in alteration in both axial and cross-stream velocities. Important to note that the extent of asymmetry which can be represented by $(\Gamma_{\max} - \Gamma_{\min})$ is much larger for high $Pe_s$ limit solution as compared with the low $Pe_s$ limit solution. This is the reason for the larger magnitude in cross-stream velocity of the droplet for high $Pe_s$ limit. In the presence of nonuniform temperature distribution, this asymmetry increases (refer to figure 11(b)) which leads to larger droplet velocities. The effect of decreasing temperature in the direction of Poiseuille flow (i.e. $\zeta = -1$) is shown in figure 11(c) for $Ma_T = 0.01$ and in figure 11(d) for $Ma_T = 0.5$. For $Ma_T = 0.01$, $\Gamma(\theta, \varphi)$ is maximum near the northwest pole and minimum near the northeast pole which is very similar to the case of $Ma_T = 0$. This kind of distribution of $\Gamma(\theta, \varphi)$ leads to motion of droplet towards the channel centerline. However, for $Ma_T = 0.5$ we obtain $\Gamma(\theta, \varphi)$ is maximum near the southeast pole and minimum near the southwest pole which results in cross-stream motion of the droplet away from the channel centerline.

## 5. Conclusions

In the present paper, we have analyzed the droplet motion in an unbounded Poiseuille flow under the combined influence of thermocapillary-induced and surfactant-induced Marangoni stresses. In the absence of fluid inertia, thermal convection and shape deformation, we have obtained asymptotic solution for the temperature field, surfactant concentration and flow field in the following two limiting conditions: (*i*) diffusion dominated surfactant transport considering $Pe_s \ll 1$, and (ii) convection dominated surfactant transport considering $Pe_s \rightarrow \infty$. Analytical expression of the velocity of a force-free droplet is obtained for both the limits. After studying effects of different parameters, the following conclusions can be made:

(*i*) Low $Pe_s$ analysis shows that when imposed temperature field increases in the direction of Poiseuille flow, a surfactant-laden droplet has both axial (along the direction of Poiseuille flow) and cross-stream (towards the flow centerline) velocity components. Magnitude of both the components increases with increase in $Ma_T$.

(*ii*) Low $Pe_s$ analysis shows that when imposed temperature field decreases in the direction of Poiseuille flow, a surfactant-laden droplet has both axial (in the direction of Poiseuille flow or opposite to it) and cross-stream (towards or away from the flow centerline) velocity components. There is a critical value of the thermal Marangoni number $Ma_{T,z}^*$ (and $Ma_{T,x}^*$) for which the axial (and cross-stream) velocity vanishes. In sharp contrast to the case of isothermal flow in which a surfactant-laden droplet always moves towards the flow centerline, a surfactant laden droplet moves away from the flow centerline for $Ma_T > Ma_{T,x}^*$ in a non-isothermal flow.

(*iii*) High $Pe_s$ analysis shows that the axial velocity of the droplet is independent of thermal effect. The axial velocity is same as the velocity of a spherical solid particle. However, the cross-stream motion is strongly dominated by the thermal conductivity ratio $(\delta)$ and $Ma_T$. When imposed temperature field increases in the direction of Poiseuille flow, the magnitude of the cross-stream velocity (towards the flow centreline) increase in $Ma_T$.

(*iv*) High $Pe_s$ analysis also shows when imposed temperature field decreases in the direction of Poiseuille flow, a surfactant-laden droplet either moves towards the flow centreline (for $Ma_T < Ma_{T,x}^*$) or moves away from the flow centreline (for $Ma_T > Ma_{T,x}^*$).

## Acknowledgements

The authors acknowledge funding from SRIC, Indian Institute of Technology Kharagpur under the Project 'Centre of Excellence for Training and Research in Microfluidics'.

## Appendix A : Velocity and pressure fields in the low $Pe_s$ limit

The leading-order velocity and pressure fields are obtained as

$$
\left.\begin{aligned}
\mathbf{u}_i^{(0)} &= \begin{bmatrix} \boldsymbol{\nabla}\times\left(\mathbf{r}\chi_1^{(0)}\right)+\boldsymbol{\nabla}\left(\Phi_1^{(0)}+\Phi_2^{(0)}+\Phi_3^{(0)}\right) \\ +\dfrac{r^2}{\lambda}\boldsymbol{\nabla}\left(\dfrac{1}{5}\,p_1^{(0)}+\dfrac{5}{42}\,p_2^{(0)}+\dfrac{1}{12}\,p_3^{(0)}\right)-\dfrac{\mathbf{r}}{\lambda}\left(\dfrac{1}{10}\,p_1^{(0)}+\dfrac{2}{21}\,p_2^{(0)}+\dfrac{1}{12}\,p_3^{(0)}\right) \end{bmatrix}, \\[2mm]
p_i^{(0)} &= p_1^{(0)}+p_2^{(0)}+p_3^{(0)}, \\[2mm]
\mathbf{u}_e^{(0)} &= \left(\mathbf{V}_\infty-\mathbf{U}^{(0)}\right)+\begin{bmatrix} \boldsymbol{\nabla}\times\left(\mathbf{r}\chi_{-2}^{(0)}\right)+\boldsymbol{\nabla}\left(\Phi_{-2}^{(0)}+\Phi_{-3}^{(0)}+\Phi_{-4}^{(0)}\right) \\ -r^2\boldsymbol{\nabla}\left(-\dfrac{1}{2}\,p_{-2}^{(0)}+\dfrac{1}{30}\,p_{-4}^{(0)}\right)+\mathbf{r}\left(2\,p_{-2}^{(0)}+\dfrac{1}{2}\,p_{-3}^{(0)}+\dfrac{4}{15}\,p_{-4}^{(0)}\right) \end{bmatrix}, \\[2mm]
p_e^{(0)} &= p_\infty+\left(p_{-2}^{(0)}+p_{-3}^{(0)}+p_{-4}^{(0)}\right).
\end{aligned}\right\} \tag{A1}
$$

The growing spherical solid harmonics present in equation (A1) are obtained as

$$
\left.\begin{aligned}
p_1^{(0)} &= \lambda r\left\{A_{1,0}^{(0)}P_{1,0}\left(\cos\theta\right)+A_{1,1}^{(0)}\cos\varphi P_{1,1}\left(\cos\theta\right)+\hat{A}_{1,1}^{(0)}\sin\varphi P_{1,1}\left(\cos\theta\right)\right\}, \\[1mm]
p_2^{(0)} &= \lambda r^2 A_{2,1}^{(0)}\cos\varphi P_{2,1}\left(\cos\theta\right),\quad p_3^{(0)} = \lambda r^3 A_{3,0}^{(0)}P_{3,0}\left(\cos\theta\right), \\[1mm]
\Phi_1^{(0)} &= r\left\{B_{1,0}^{(0)}P_{1,0}\left(\cos\theta\right)+B_{1,1}^{(0)}\cos\varphi P_{1,1}\left(\cos\theta\right)+\hat{B}_{1,1}^{(0)}\sin\varphi P_{1,1}\left(\cos\theta\right)\right\}, \\[1mm]
\Phi_2^{(0)} &= r^2 B_{2,1}^{(0)}\cos\varphi P_{2,1}\left(\cos\theta\right),\quad \Phi_3^{(0)} = r^3 B_{3,0}^{(0)}P_{3,0}\left(\cos\theta\right), \\[1mm]
\chi_1^{(0)} &= r\hat{C}_{1,1}^{(0)}\sin\varphi P_{1,1}\left(\cos\theta\right),
\end{aligned}\right\} \tag{A2}
$$

where the unknown coefficients are obtained as

$$A_{1,0}^{(0)} = -\frac{5\left(2Ma_T T_{1,0}^{(0)} - 15\alpha_{1,0}^{(0)} - 3\beta_{1,0}^{(0)}\right)}{3\lambda+3}, \; B_{1,0}^{(0)} = \frac{2Ma_T T_{1,0}^{(0)} - 15\alpha_{1,0}^{(0)} - 3\beta_{1,0}^{(0)}}{2\left(3\lambda+3\right)}, \; A_{2,1}^{(0)} = \frac{105\beta_{2,1}^{(0)}}{2\left(5\lambda+5\right)},$$

$$B_{2,1}^{(0)} = -\frac{15\beta_{2,1}^{(0)}}{4\left(5\lambda+5\right)}, \; A_{3,0}^{(0)} = \frac{15\beta_{3,0}^{(0)}}{7\left(\lambda+1\right)}, \; B_{3,0}^{(0)} = -\frac{5\beta_{3,0}^{(0)}}{6\left(\lambda+1\right)}, \; A_{1,1}^{(0)} = \frac{5\beta_{1,1}^{(0)}}{\left(\lambda+1\right)}, \; B_{1,1}^{(0)} = -\frac{\beta_{1,1}^{(0)}}{2\left(\lambda+1\right)},$$

$$\hat{A}_{1,1}^{(0)} = \frac{5}{\left(\lambda+1\right)}\hat{\beta}_{1,1}^{(0)}, \; \hat{B}_{1,1}^{(0)} = -\frac{\hat{\beta}_{1,1}^{(0)}}{2\left(\lambda+1\right)}, \; \hat{C}_{1,1}^{(0)} = \frac{1}{2}\hat{\gamma}_{1,1}^{(0)},$$

$$\tag{A3}$$

with

$$T_{1,0}^{(0)} = \frac{3\zeta}{\left(\delta+2\right)}, \;\; \alpha_{1,0}^{(0)} = -\frac{2}{5R^2}, \;\; \beta_{1,0}^{(0)} = 1 - \left(\frac{e}{R}\right)^2 - U_z^{(0)},$$

$$\beta_{1,1}^{(0)} = -U_x^{(0)}, \;\; \hat{\beta}_{1,1}^{(0)} = -U_y^{(0)}, \;\; \beta_{2,1}^{(0)} = -\frac{2}{3}\frac{e}{R^2},$$

$$\beta_{3,0}^{(0)} = \frac{2}{5R^2}, \;\; \hat{\gamma}_{1,1}^{(0)} = \frac{2e}{R^2}.$$

$$\tag{A4}$$

The decaying spherical solid harmonics present in equation (A1) are obtained as

$$p_{-2}^{(0)} = r^{-2}\left\{A_{-2,0}^{(0)}P_{1,0}\left(\cos\theta\right) + A_{-2,1}^{(0)}\cos\varphi P_{1,1}\left(\cos\theta\right) + \hat{A}_{-2,1}^{(0)}\sin\varphi P_{1,1}\left(\cos\theta\right)\right\},$$

$$p_{-3}^{(0)} = r^{-3}A_{-3,1}^{(0)}\cos\varphi P_{2,1}\left(\cos\theta\right), \; p_{-4}^{(0)} = r^{-4}A_{-4,0}^{(0)}P_{3,0}\left(\cos\theta\right),$$

$$\Phi_{-2}^{(0)} = r^{-2}\left\{B_{-2,0}^{(0)}P_{1,0}\left(\cos\theta\right) + B_{-2,1}^{(0)}\cos\varphi P_{1,1}\left(\cos\theta\right) + \hat{B}_{-2,1}^{(0)}\sin\varphi P_{1,1}\left(\cos\theta\right)\right\},$$

$$\Phi_{-3}^{(0)} = r^{-3}B_{-3,1}^{(0)}\cos\varphi P_{2,1}\left(\cos\theta\right), \;\; \Phi_{-4}^{(0)} = r^{-4}B_{-4,0}^{(0)}P_{3,0}\left(\cos\theta\right),$$

$$\tag{A5}$$

where the unknown coefficients are obtained as

$$A_{-2,0}^{(0)} = -\frac{2Ma_T T_{1,0}^{(0)} + 15\lambda\alpha_{1,0}^{(0)} + 9\lambda\beta_{1,0}^{(0)} + 6\beta_{1,0}^{(0)}}{6\left(\lambda+1\right)}, \;\; A_{-2,1}^{(0)} = -\frac{\left(2+3\lambda\right)}{2\left(1+\lambda\right)}\beta_{1,1}^{(0)}, \;\; \hat{A}_{-2,1}^{(0)} = -\frac{\left(2+3\lambda\right)}{2\left(1+\lambda\right)}\hat{\beta}_{1,1}^{(0)},$$

$$B_{-2,0}^{(0)} = -\frac{2Ma_T T_{1,0}^{(0)} + 9\lambda\alpha_{1,0}^{(0)} - 6\alpha_{1,0}^{(0)} + 3\lambda\beta_{1,0}^{(0)}}{12\left(\lambda+1\right)}, \;\; \hat{B}_{-2,1}^{(0)} = -\frac{\lambda}{4\left(\lambda+1\right)}\hat{\beta}_{1,1}^{(0)}, \; B_{-2,1}^{(0)} = -\frac{\lambda}{4\left(\lambda+1\right)}\beta_{1,1}^{(0)},$$

$$A_{-3,1}^{(0)} = -\frac{5\lambda\beta_{2,1}^{(0)} + 2\beta_{2,1}^{(0)}}{\left(\lambda+1\right)}, B_{-3,1}^{(0)} = \frac{-3\lambda\beta_{2,1}^{(0)}}{6\left(\lambda+1\right)}, A_{-4,0}^{(0)} = -\frac{35\lambda\beta_{3,0}^{(0)} + 10\beta_{3,0}^{(0)}}{4\left(\lambda+1\right)}, B_{-4,0}^{(0)} = \frac{-5\lambda\beta_{3,0}^{(0)}}{8\left(\lambda+1\right)}.$$

$$\tag{A6}$$

The far-field pressure field is of the form

$$p_\infty = 10r\alpha_{1,0}^{(0)}P_{1,0}\left(\cos\theta\right) + 7r^2\alpha_{2,1}^{(0)}\cos\varphi P_{2,1}\left(\cos\theta\right) + 6r^3\alpha_{3,0}^{(0)}P_{3,0}\left(\cos\theta\right). \tag{A7}$$

The $O\left(Pe_s\right)$ velocity and pressure fields are obtained as

$$\mathbf{u}_i^{(Pe_s)} = \begin{bmatrix} \boldsymbol{\nabla}\left(\Phi_1^{(Pe_s)} + \Phi_2^{(Pe_s)} + \Phi_3^{(Pe_s)}\right) \\ + \dfrac{r^2}{\lambda}\boldsymbol{\nabla}\left(\dfrac{p_1^{(Pe_s)}}{5} + \dfrac{5p_2^{(Pe_s)}}{42} + \dfrac{p_3^{(Pe_s)}}{12}\right) - \dfrac{\mathbf{r}}{\lambda}\left(\dfrac{p_1^{(Pe_s)}}{10} + \dfrac{2p_2^{(Pe_s)}}{21} + \dfrac{p_3^{(Pe_s)}}{12}\right) \end{bmatrix},$$

$$p_i^{(Pe_s)} = p_1^{(Pe_s)} + p_2^{(Pe_s)} + p_3^{(Pe_s)},$$

$$\mathbf{u}_e^{(Pe_s)} = -\mathbf{U}^{(Pe_s)} + \sum_{n=1}^{\infty}\begin{bmatrix} \boldsymbol{\nabla}\left(\Phi_{-2}^{(Pe_s)} + \Phi_{-3}^{(Pe_s)} + \Phi_{-4}^{(Pe_s)}\right) \\ -r^2\boldsymbol{\nabla}\left(-\dfrac{p_{-2}^{(Pe_s)}}{2} + \dfrac{p_{-4}^{(Pe_s)}}{30}\right) + \mathbf{r}\left(2p_{-2}^{(Pe_s)} + \dfrac{p_{-3}^{(Pe_s)}}{2} + \dfrac{4p_{-4}^{(Pe_s)}}{15}\right) \end{bmatrix},$$

$$p_e^{(Pe_s)} = p_{-2}^{(Pe_s)} + p_{-3}^{(Pe_s)} + p_{-4}^{(Pe_s)}.$$

(A8)

The growing spherical solid harmonics present in equation (A8) are obtained as

$$p_1^{(Pe_s)} = \lambda r\left\{A_{1,0}^{(Pe_s)}P_{1,0}\left(\cos\theta\right) + A_{1,1}^{(Pe_s)}\cos\varphi P_{1,1}\left(\cos\theta\right) + \hat{A}_{1,1}^{(Pe_s)}\sin\varphi P_{1,1}\left(\cos\theta\right)\right\},$$

$$p_2^{(Pe_s)} = \lambda r^2 A_{2,1}^{(Pe_s)}\cos\varphi P_{2,1}\left(\cos\theta\right), \quad p_3^{(Pe_s)} = \lambda r^3 A_{3,0}^{(Pe_s)}P_{3,0}\left(\cos\theta\right),$$

$$\Phi_1^{(Pe_s)} = r\left\{B_{1,0}^{(Pe_s)}P_{1,0}\left(\cos\theta\right) + B_{1,1}^{(Pe_s)}\cos\varphi P_{1,1}\left(\cos\theta\right) + \hat{B}_{1,1}^{(Pe_s)}\sin\varphi P_{1,1}\left(\cos\theta\right)\right\},$$

$$\Phi_2^{(Pe_s)} = r^2 B_{2,1}^{(Pe_s)}\cos\varphi P_{2,1}\left(\cos\theta\right), \quad \Phi_3^{(Pe_s)} = r^3 B_{3,0}^{(Pe_s)}P_{3,0}\left(\cos\theta\right),$$

(A9)

where the unknown coefficients are obtained as

$$A_{1,0}^{(Pe_s)} = -\frac{5\left\{2Ma_{\Gamma}\Gamma_{1,0}^{(Pe_s)} - 3\beta_{1,0}^{(Pe_s)}\right\}}{3\left(\lambda+1\right)}, B_{1,0}^{(Pe_s)} = \frac{2Ma_{\Gamma}\Gamma_{1,0}^{(Pe_s)} - 3\beta_{1,0}^{(Pe_s)}}{2\left(3\lambda+3\right)},$$

$$A_{1,1}^{(Pe_s)} = \frac{5\beta_{1,1}^{(Pe_s)}}{\left(\lambda+1\right)}, B_{1,1}^{(Pe_s)} = -\frac{\beta_{1,1}^{(Pe_s)}}{2\left(\lambda+1\right)}, \hat{A}_{1,1}^{(Pe_s)} = \frac{5}{\left(\lambda+1\right)}\hat{\beta}_{1,1}^{(Pe_s)}, \hat{B}_{1,1}^{(Pe_s)} = -\frac{\hat{\beta}_{1,1}^{(Pe_s)}}{2\left(\lambda+1\right)},$$

$$A_{2,1}^{(Pe_s)} = -\frac{21Ma_{\Gamma}\Gamma_{2,1}^{(Pe_s)}}{5\left(\lambda+1\right)}, B_{2,1}^{(Pe_s)} = \frac{3Ma_{\Gamma}\Gamma_{2,1}^{(Pe_s)}}{10\left(\lambda+1\right)},$$

$$A_{3,0}^{(Pe_s)} = -\frac{36Ma_{\Gamma}\Gamma_{3,0}^{(Pe_s)}}{7\left(\lambda+1\right)}, B_{3,0}^{(Pe_s)} = \frac{2Ma_{\Gamma}\Gamma_{3,0}^{(Pe_s)}}{7\left(\lambda+1\right)},$$

(A10)

with $\beta_{1,0}^{(Pe_s)} = -U_z^{(Pe_s)}$, $\beta_{1,1}^{(Pe_s)} = -U_x^{(Pe_s)}$ and $\hat{\beta}_{1,1}^{(Pe_s)} = -U_y^{(Pe_s)}$.

The growing spherical solid harmonics present in equation (A8) are obtained as

$$p_{-2}^{(Pe_s)} = r^{-2}\left\{A_{-2,0}^{(Pe_s)}P_{1,0}\left(\cos\theta\right) + A_{-2,1}^{(Pe_s)}\cos\varphi P_{1,1}\left(\cos\theta\right) + \hat{A}_{-2,1}^{(Pe_s)}\sin\varphi P_{1,1}\left(\cos\theta\right)\right\},$$

$$p_{-3}^{(Pe_s)} = r^{-3}A_{-3,1}^{(Pe_s)}\cos\varphi P_{2,1}\left(\cos\theta\right), \quad p_{-4}^{(Pe_s)} = r^{-4}A_{-4,0}^{(Pe_s)}P_{3,0}\left(\cos\theta\right),$$

$$\Phi_{-2}^{(Pe_s)} = r^{-2}\left\{B_{-2,0}^{(Pe_s)}P_{1,0}\left(\cos\theta\right) + B_{-2,1}^{(Pe_s)}\cos\varphi P_{1,1}\left(\cos\theta\right) + \hat{B}_{-2,1}^{(Pe_s)}\sin\varphi P_{1,1}\left(\cos\theta\right)\right\},$$

$$\Phi_{-3}^{(0)} = r^{-3}B_{-3,1}^{(Pe_s)}\cos\varphi P_{2,1}\left(\cos\theta\right), \quad \Phi_{-4}^{(Pe_s)} = r^{-4}B_{-4,0}^{(Pe_s)}P_{3,0}\left(\cos\theta\right),$$

(A11)

where the unknown coefficients are obtained as

$$A_{-2,0}^{(Pe_s)} = -\frac{9\lambda\beta_{1,0}^{(Pe_s)} + 2Ma_{\Gamma}\Gamma_{1,0}^{(Pe_s)} + 6\beta_{1,0}^{(Pe_s)}}{6(\lambda+1)}, A_{-2,1}^{(Pe_s)} = -\frac{(2+3\lambda)}{2(1+\lambda)}\beta_{1,1}^{(Pe_s)}, \quad \hat{A}_{-2,1}^{(0)} = -\frac{(2+3\lambda)}{2(1+\lambda)}\hat{\beta}_{1,1}^{(Pe_s)},$$

$$B_{-2,0}^{(Pe_s)} = -\frac{2Ma_{\Gamma}\Gamma_{1,0}^{(Pe_s)} + 3\lambda\beta_{1,0}^{(Pe_s)}}{12(\lambda+1)}, \quad \hat{B}_{-2,1}^{(Pe_s)} = -\frac{\lambda}{4(\lambda+1)}\hat{\beta}_{1,1}^{(Pe_s)}, \quad B_{-2,1}^{(Pe_s)} = -\frac{\lambda}{4(\lambda+1)}\beta_{1,1}^{(Pe_s)},$$

$$A_{-3,1}^{(Pe_s)} = -\frac{6Ma_{\Gamma}\Gamma_{2,1}^{(Pe_s)}}{5(\lambda+1)}, \quad B_{-3,1}^{(Pe_s)} = -\frac{Ma_{\Gamma}\Gamma_{2,1}^{(Pe_s)}}{5(\lambda+1)}, A_{-4,0}^{(Pe_s)} = -\frac{15Ma_{\Gamma}\Gamma_{3,0}^{(Pe_s)}}{7(\lambda+1)}, B_{-4,0}^{(Pe_s)} = -\frac{3Ma_{\Gamma}\Gamma_{3,0}^{(Pe_s)}}{14(\lambda+1)}.$$

$$(A12)$$

The $O\left(Pe_s^2\right)$ velocity and pressure fields are obtained as

$$\mathbf{u}_i^{(Pe_s^2)} = \begin{bmatrix} \boldsymbol{\nabla}\left(\Phi_1^{(Pe_s^2)} + \Phi_2^{(Pe_s^2)} + \Phi_3^{(Pe_s^2)}\right) \\ + \frac{r^2}{\lambda}\boldsymbol{\nabla}\left(\frac{1}{5}p_1^{(Pe_s^2)} + \frac{5}{42}p_2^{(Pe_s^2)} + \frac{1}{12}p_3^{(Pe_s^2)}\right) - \frac{\mathbf{r}}{\lambda}\left(\frac{1}{10}p_1^{(Pe_s^2)} + \frac{2}{21}p_2^{(Pe_s^2)} + \frac{1}{12}p_3^{(Pe_s^2)}\right) \end{bmatrix},$$

$$p_i^{(Pe_s^2)} = p_1^{(Pe_s^2)} + p_2^{(Pe_s^2)} + p_3^{(Pe_s^2)},$$

$$\mathbf{u}_e^{(Pe_s^2)} = -\mathbf{U}^{(Pe_s^2)} + \sum_{n=1}^{\infty}\begin{bmatrix} \boldsymbol{\nabla}\left(\Phi_{-2}^{(Pe_s^2)} + \Phi_{-3}^{(Pe_s^2)} + \Phi_{-4}^{(Pe_s^2)}\right) \\ -r^2\boldsymbol{\nabla}\left(-\frac{1}{2}p_{-2}^{(Pe_s^2)} + \frac{1}{30}p_{-4}^{(Pe_s^2)}\right) + \mathbf{r}\left(2p_{-2}^{(Pe_s^2)} + \frac{1}{2}p_{-3}^{(Pe_s^2)} + \frac{4}{15}p_{-4}^{(Pe_s^2)}\right) \end{bmatrix},$$

$$p_e^{(Pe_s^2)} = p_{-2}^{(Pe_s^2)} + p_{-3}^{(Pe_s^2)} + p_{-4}^{(Pe_s^2)}.$$

$$(A13)$$

The growing spherical solid harmonics present in equation (A13) are obtained as

$$p_1^{(Pe_s^2)} = \lambda r\left\{A_{1,0}^{(Pe_s^2)}P_{1,0}(\cos\theta) + A_{1,1}^{(Pe_s^2)}\cos\varphi P_{1,1}(\cos\theta) + \hat{A}_{1,1}^{(Pe_s^2)}\sin\varphi P_{1,1}(\cos\theta)\right\},$$

$$p_2^{(Pe_s^2)} = \lambda r^2 A_{2,1}^{(Pe_s^2)}\cos\varphi P_{2,1}(\cos\theta), p_3^{(Pe_s^2)} = \lambda r^3\left\{A_{3,0}^{(Pe_s^2)}P_{3,0}(\cos\theta) + A_{3,1}^{(Pe_s^2)}\cos\varphi P_{3,1}(\cos\theta)\right\},$$

$$\Phi_1^{(Pe_s^2)} = r\left\{B_{1,0}^{(Pe_s^2)}P_{1,0}(\cos\theta) + B_{1,1}^{(Pe_s^2)}\cos\varphi P_{1,1}(\cos\theta) + \hat{B}_{1,1}^{(Pe_s^2)}\sin\varphi P_{1,1}(\cos\theta)\right\},$$

$$\Phi_2^{(Pe_s^2)} = r^2 B_{2,1}^{(Pe_s^2)}\cos\varphi P_{2,1}(\cos\theta), \Phi_3^{(Pe_s^2)} = r^3\left\{B_{3,0}^{(Pe_s)}P_{3,0}(\cos\theta) + B_{3,1}^{(Pe_s^2)}\cos\varphi P_{3,1}(\cos\theta)\right\},$$

$$(A14)$$

where the unknown coefficients are obtained as

$$A_{1,0}^{(Pe_s^2)} = -\frac{5\left(2Ma_{\Gamma}\Gamma_{1,0}^{(Pe_s^2)} - 3\beta_{1,0}^{(Pe_s^2)}\right)}{3(\lambda+1)}, \quad B_{1,0}^{(Pe_s^2)} = \frac{2Ma_{\Gamma}\Gamma_{1,0}^{(Pe_s^2)} - 3\beta_{1,0}^{(Pe_s^2)}}{6(\lambda+1)},$$

$$A_{1,1}^{(Pe_s^2)} = -\frac{5\left(2Ma_{\Gamma}\Gamma_{1,1}^{(Pe_s^2)} - 3\beta_{1,1}^{(Pe_s^2)}\right)}{3(\lambda+1)}, B_{1,1}^{(Pe_s)} = \frac{2Ma_{\Gamma}\Gamma_{1,1}^{(Pe_s^2)} - 3\beta_{1,1}^{(Pe_s^2)}}{6(\lambda+1)},$$

$$(A15)$$

$$\left.\begin{aligned}
\hat{A}_{1,1}^{(Pe_s)} &= \frac{5}{(\lambda+1)}\hat{\beta}_{1,1}^{(Pe_s)}, \hat{B}_{1,1}^{(Pe_s)} = -\frac{\hat{\beta}_{1,1}^{(Pe_s)}}{2(\lambda+1)}, \ A_{2,1}^{(Pe_s^2)} = -\frac{21Ma_\Gamma\Gamma_{2,1}^{(Pe_s^2)}}{5(\lambda+1)}, \\
B_{2,1}^{(Pe_s^2)} &= \frac{3Ma_\Gamma\Gamma_{2,1}^{(Pe_s^2)}}{10(\lambda+1)}, A_{3,0}^{(Pe_s^2)} = -\frac{36Ma_\Gamma\Gamma_{3,0}^{(Pe_s^2)}}{7(\lambda+1)}, B_{3,0}^{(Pe_s^2)} = \frac{2Ma_\Gamma\Gamma_{3,0}^{(Pe_s^2)}}{7(\lambda+1)}, \\
A_{3,1}^{(Pe_s^2)} &= -\frac{36Ma_\Gamma\Gamma_{3,1}^{(Pe_s^2)}}{7(\lambda+1)}, B_{3,1}^{(Pe_s^2)} = \frac{2Ma_\Gamma\Gamma_{3,1}^{(Pe_s^2)}}{7(\lambda+1)},
\end{aligned}\right\} \quad \text{(A16)}$$

with $\beta_{1,0}^{(Pe_s^2)} = -U_z^{(Pe_s^2)}$, $\beta_{1,1}^{(Pe_s^2)} = -U_x^{(Pe_s^2)}$ and $\hat{\beta}_{1,1}^{(Pe_s^2)} = -U_y^{(Pe_s^2)}$. The different constants present in the expression of surfactant concentration $\left(\Gamma_{n,m}^{(Pe_s^2)}\right)$ for $O(Pe_s^2)$ are given in Appendix B. The decaying spherical solid harmonics present in equation (A13) are obtained as

$$\left.\begin{aligned}
p_{-2}^{(Pe_s^2)} &= r^{-2}\left\{A_{-2,0}^{(Pe_s^2)}P_{1,0}(\cos\theta) + A_{-2,1}^{(Pe_s^2)}\cos\varphi P_{1,1}(\cos\theta) + \hat{A}_{-2,1}^{(Pe_s^2)}\sin\varphi P_{1,1}(\cos\theta)\right\}, \\
p_{-3}^{(Pe_s^2)} &= r^{-3}A_{-3,1}^{(Pe_s^2)}\cos\varphi P_{2,1}(\cos\theta), p_{-4}^{(Pe_s^2)} = r^{-4}\left\{A_{-4,0}^{(Pe_s^2)}P_{3,0}(\cos\theta) + A_{-4,1}^{(Pe_s^2)}\cos\varphi P_{3,1}(\cos\theta)\right\}, \\
\Phi_{-2}^{(Pe_s^2)} &= r^{-2}\left\{B_{-2,0}^{(Pe_s^2)}P_{1,0}(\cos\theta) + B_{-2,1}^{(Pe_s^2)}\cos\varphi P_{1,1}(\cos\theta) + \hat{B}_{-2,1}^{(Pe_s^2)}\sin\varphi P_{1,1}(\cos\theta)\right\}, \\
\Phi_{-3}^{(0)} &= r^{-3}B_{-3,1}^{(Pe_s^2)}\cos\varphi P_{2,1}(\cos\theta), \Phi_{-4}^{(Pe_s^2)} = r^{-4}\left\{B_{-4,0}^{(Pe_s^2)}P_{3,0}(\cos\theta) + B_{-4,1}^{(Pe_s^2)}\cos\varphi P_{3,1}(\cos\theta)\right\},
\end{aligned}\right\} \text{(A17)}$$

where the unknown coefficients are obtained as

$$\left.\begin{aligned}
A_{-2,0}^{(Pe_s^2)} &= -\frac{9\lambda\beta_{1,0}^{(Pe_s^2)} + 2Ma_\Gamma\Gamma_{1,0}^{(Pe_s^2)} + 6\beta_{1,0}^{(Pe_s^2)}}{6(\lambda+1)}, \ A_{-2,1}^{(Pe_s^2)} = -\left\{\frac{Ma_\Gamma\Gamma_{1,1}^{(Pe_s^2)}}{3(\lambda+1)} + \frac{3\lambda+2}{2(\lambda+1)}\beta_{1,1}^{(Pe_s^2)}\right\} \\
\hat{A}_{-2,1}^{(Pe_s^2)} &= -\frac{3\lambda+2}{2(\lambda+1)}\hat{\beta}_{1,1}^{(Pe_s^2)}, B_{-2,1}^{(Pe_s^2)} = -\left\{\frac{Ma_\Gamma\Gamma_{1,1}^{(Pe_s^2)}}{6(\lambda+1)} + \frac{\lambda}{4(\lambda+1)}\beta_{1,1}^{(Pe_s^2)}\right\}, \ \hat{B}_{-2,1}^{(Pe_s^2)} = -\frac{\lambda}{4(\lambda+1)}\hat{\beta}_{1,1}^{(Pe_s^2)}, \\
B_{-2,0}^{(Pe_s^2)} &= -\frac{2Ma_\Gamma\Gamma_{1,0}^{(Pe_s^2)} + 3\lambda\beta_{1,0}^{(Pe_s^2)}}{12(\lambda+1)}, A_{-3,1}^{(Pe_s^2)} = -\frac{6Ma_\Gamma\Gamma_{2,1}^{(Pe_s^2)}}{5(\lambda+1)}, \ B_{-3,1}^{(Pe_s^2)} = -\frac{Ma_\Gamma\Gamma_{2,1}^{(Pe_s^2)}}{5(\lambda+1)}, \\
A_{-4,1}^{(Pe_s^2)} &= -\frac{15Ma_\Gamma\Gamma_{3,1}^{(Pe_s^2)}}{7(\lambda+1)}, \ B_{-4,1}^{(Pe_s^2)} = -\frac{3Ma_\Gamma\Gamma_{3,1}^{(Pe_s^2)}}{14(\lambda+1)}, \ A_{-4,0}^{(Pe_s^2)} = -\frac{15Ma_\Gamma\Gamma_{3,0}^{(Pe_s^2)}}{7(\lambda+1)}, B_{-4,0}^{(Pe_s^2)} = -\frac{3Ma_\Gamma\Gamma_{3,0}^{(Pe_s^2)}}{14(\lambda+1)}.
\end{aligned}\right\} \text{(A18)}$$

## Appendix B : Expressions for the constant coefficients present in equation (39) for surfactant concentration of $O(Pe_s^2)$

The coefficients present in the $O\left(Pe_s^2\right)$ surface harmonics of surfactant concentration are obtained as

$$
\left.\begin{aligned}
\Gamma_{1,0}^{\left(Pe_s^2\right)} &= \frac{\left(2\delta + 3Ma_T\zeta R^2 + 4\right)Ma_\Gamma}{\left(9\delta\lambda^2 + 12\delta\lambda + 4\delta + 18\lambda^2 + 8 + 24\lambda\right)R^2}, \\[2mm]
\Gamma_{1,1}^{\left(Pe_s^2\right)} &= -\frac{\left(\begin{array}{l}70\delta\lambda^2 + 109\delta\lambda + 40\delta + 140\lambda^2 + 218\lambda + 80 \\ +105Ma_T\zeta R^2\lambda^2 + 168Ma_T\zeta R^2\lambda + 63Ma_T\zeta R^2\end{array}\right)e}{70\left(7\delta\lambda + 8\delta\lambda^2 + 3\delta\lambda^3 + 2\delta + 6\lambda^3 + 16\lambda^2 + 4 + 14\lambda\right)R^4}, \\[2mm]
\Gamma_{2,0}^{\left(Pe_s^2\right)} &= \frac{\left[\begin{array}{l}1134\left(\lambda+1\right)^2 Ma_T^2\zeta^2 R^4 + 27\left(47\lambda + 50\right)\left(\lambda+1\right)\left(\delta+2\right)Ma_T\zeta R^2 \\ +\left(\delta+2\right)^2\left(567e^2\lambda^3 + 750\lambda + 1404e^2\lambda^2 + 288e^2 + 400 + 351\lambda^2 + 1116e^2\lambda\right)\end{array}\right]}{378R^4\left(\lambda+1\right)^2\left(3\lambda+2\right)^2\left(\delta+2\right)^2}, \\[2mm]
\Gamma_{2,1}^{\left(Pe_s^2\right)} &= \frac{1}{15}\frac{Ma_\Gamma\zeta e}{\left(\lambda^2 + 2\lambda + 1\right)R^2}, \quad \Gamma_{2,2}^{\left(Pe_s^2\right)} = -\frac{1}{252}\frac{\left(7\lambda+4\right)e^2}{\left(\lambda^2 + 2\lambda + 1\right)R^4}, \Gamma_{3,0}^{\left(Pe_s^2\right)} = -\frac{1}{42}\frac{Ma_\Gamma}{R^2\left(\lambda^2 + 2\lambda + 1\right)}, \\[2mm]
\Gamma_{3,1}^{\left(Pe_s^2\right)} &= \frac{\left[432\left(\lambda+1\right)Ma_T\zeta R^2 + \left(45\lambda^2 + 351\lambda + 310\right)\left(\delta+2\right)\right]e}{1080\left(3\lambda+2\right)\left(\lambda+1\right)^2\left(\delta+2\right)R^4}.
\end{aligned}\right\}
\tag{B1}
$$

### Appendix C: Velocity and pressure fields in the high $Pe_s$ limit

The leading-order velocity and pressure fields are obtained as

$$
\left.\begin{aligned}
\mathbf{u}_i^{(0)} &= \nabla\times\left(\mathbf{r}\chi_1^{(0)}\right), \\
p_i^{(0)} &= 0, \\
\mathbf{u}_e^{(0)} &= \left(\mathbf{V}_\infty - \mathbf{U}^{(0)}\right) + \sum_{n=1}^{\infty}\left[\begin{array}{l}\nabla\left(\Phi_{-2}^{(0)} + \Phi_{-3}^{(0)} + \Phi_{-4}^{(0)}\right) \\ -r^2\nabla\left(-\dfrac{p_{-2}^{(0)}}{2} + \dfrac{p_{-4}^{(0)}}{30}\right) + \mathbf{r}\left(2p_{-2}^{(0)} + \dfrac{p_{-3}^{(0)}}{2} + \dfrac{4p_{-4}^{(0)}}{15}\right)\end{array}\right], \\
p_e^{(0)} &= p_\infty + \left(p_{-2}^{(0)} + p_{-3}^{(0)} + p_{-4}^{(0)}\right).
\end{aligned}\right\}
\tag{C1}
$$

The growing spherical solid harmonics present in equation (C1) are obtained as

$$
\chi_1^{(0)} = \frac{1}{2}\hat{\gamma}_{1,1}^{(0)}r\sin\varphi P_{1,1}\left(\cos\theta\right).
\tag{C2}
$$

The decaying spherical solid harmonics present in equation (C1) are obtained as

$$p_{-2}^{(0)} = r^{-2}\left\{A_{-2,0}^{(0)}P_{1,0}\left(\cos\theta\right) + A_{-2,1}^{(0)}\cos\varphi P_{1,1}\left(\cos\theta\right) + \hat{A}_{-2,1}^{(0)}\sin\varphi P_{1,1}\left(\cos\theta\right)\right\},$$

$$p_{-3}^{(0)} = r^{-3}A_{-3,1}^{(0)}\cos\varphi P_{2,1}\left(\cos\theta\right), \quad p_{-4}^{(0)} = r^{-4}A_{-4,0}^{(0)}P_{3,0}\left(\cos\theta\right),$$

$$\Phi_{-2}^{(0)} = r^{-2}\left\{B_{-2,0}^{(0)}P_{1,0}\left(\cos\theta\right) + B_{-2,1}^{(0)}\cos\varphi P_{1,1}\left(\cos\theta\right) + \hat{B}_{-2,1}^{(0)}\sin\varphi P_{1,1}\left(\cos\theta\right)\right\},$$

$$\Phi_{-3}^{(0)} = r^{-3}B_{-3,1}^{(0)}\cos\varphi P_{2,1}\left(\cos\theta\right), \quad \Phi_{-4}^{(0)} = r^{-4}B_{-4,0}^{(0)}P_{3,0}\left(\cos\theta\right),$$

(C3)

where the unknown coefficients are obtained as

$$A_{-2,0}^{(0)} = -\frac{5}{2}\alpha_{1,0}^{(0)} - \frac{3}{2}\beta_{1,0}^{(0)}, \quad A_{-2,1}^{(0)} = -\frac{3}{2}\beta_{1,1}^{(0)}, \quad \hat{A}_{-2,1}^{(0)} = -\frac{3}{2}\hat{\beta}_{1,1}^{(0)}, \quad B_{-2,0}^{(0)} = -\frac{3}{4}\alpha_{1,0}^{(0)} - \frac{1}{4}\beta_{1,0}^{(0)},$$

$$B_{-2,1}^{(0)} = -\frac{1}{4}\beta_{1,1}^{(0)}, B_{-2,1}^{(0)} = -\frac{1}{4}\beta_{1,1}^{(0)}, A_{-3,1}^{(0)} = -5\beta_{2,1}^{(0)}, B_{-3,1}^{(0)} = -\frac{1}{2}\beta_{2,1}^{(0)}, A_{-4,0}^{(0)} = -\frac{35}{4}\beta_{3,0}^{(0)}, B_{-4,0}^{(0)} = \frac{5}{8}\beta_{3,0}^{(0)}.$$

(C4)